\documentclass{amsart}
\usepackage{amsmath}
\usepackage{amssymb}
\usepackage{amsfonts}

\newtheorem{theorem}{Theorem}

\newtheorem{corollary}[theorem]{Corollary}
\newtheorem{criterion}[theorem]{Criterion}
\newtheorem{definition}[theorem]{Definition}

\newtheorem{lemma}[theorem]{Lemma}

\newtheorem{proposition}[theorem]{Proposition}
\newtheorem{remark}[theorem]{Remark}

\newcommand{\cA}{{\mathcal A}}
\newcommand{\cB}{{\mathcal B}}
\newcommand{\C}{{\mathcal C}}
\newcommand{\cP}{{\mathcal P}}

\newcommand{\cF}{{\mathcal F}}

\newcommand{\cK}{{\mathcal K}}
\newcommand{\cS}{{\mathcal S}}
\newcommand{\cH}{{\mathcal H}}

\newcommand{\bC}{{\mathbb{C}}}
\newcommand{\bN}{{\mathbb{N}}}
\newcommand{\Cs}{{{$\hbox{\bf C}^*$}}}
\begin{document}

\title{Characterization of PPT states and measures of entanglement}

\author{W{\l}adys{\l}aw A. Majewski}
\address{Institute of Theoretical Physics and Astrophysics, Gda{\'n}sk
University, Wita Stwo\-sza~57, 80-952 Gda{\'n}sk, Poland} 
\email{fizwam@univ.gda.pl}

\author{Takashi Matsuoka}
\address{Faculty of Management of Administration and Information,
Tokyo University of Science, Suwa; 
Toyohira 5001, Chino City,
Nagano 391-0292, Japan} 
\email{matsuoka@rs.suwa.tus.ac.jp}

\author{Masanori Ohya}
\address{Department of Information Science,
Tokyo University of Science,
Yamazaki 2641, Noda City,
Chiba 278-8510, Japan} 
\email{ohya@rs.noda.tus.ac.jp}
\begin{abstract}
A detailed characterization of PPT states, both in the Heisenberg and in the Schr{\"o}dinger picture, is given.
Measures of entanglement are defined and discussed in details. Illustrative examples are provided.
\end{abstract}
\maketitle
\section{Preliminaries}

In this section we compile some basic facts on the theory of positive maps on \Cs-algebras.
To begin with, let $\cA$ and $\cB$ be \Cs-algebras (with unit), $\cA_h = \{ a \in \cA;
a = a^* \}$ - the set of all selfadjoint elements in $\cA$, $\cA^+ = \{ a \in \cA_h; a \ge 0 \}$ - the set of all positive elements in 
$\cA$, and $\cS(\cA)$ the set of all states on $\cA$, i.e. the set of all linear functionals $\varphi$
on $\cA$ such that $\varphi(1) = 1$ and $\varphi(a)\geq0$ for any $a \in \cA^+$.
In particular
$$ (\cA_h, \cA^+)\quad is \quad an \quad ordered \quad Banach \quad space. $$
We say that a linear map $\alpha : \cA \to \cB$ is positive if $\alpha(\cA^+) \subset \cB^+$.

The theory of positive maps on non-commutative algebras can be viewed as a 
jig-saw-puzzle with 
pieces whose exact form is not well known. On the other hand, as we address this paper to 
a readership interested in quantum mechanics and quantum information theory, in this section, 
we will focus our attention on some carefully selected basic concepts and fundamental results in order to  
facilitate access to main problems of that theory.
Furthermore, the relations between the theory of positive maps 
and the entanglement problem will be indicated.

\smallskip

We begin with a very strong notion of positivity:
the so called complete positivity (CP).
Namely, a linear map $\tau : \cA \to \cB$ is CP iff
\begin{equation}
\label{CP}
\tau_n : M_n(\cA) \to M_n(\cB); [a_{ij}] \mapsto [\tau(a_{ij})]
\end{equation}
is positive for all $n$. Here, $M_n(\cA)$ stands for $n \times n$ matrices with entries in $\cA$.

To explain the basic motivation for that concept we need the following notion: 
{\it an operator state
of \Cs-algebra $\cA$ on a Hilbert space $\cK$, is a CP map $\tau : \cA \to \cB(\cK)$}, where $\cB(\cK)$ stands for the set of all bounded linear operators on $\mathcal K$.
Having this concept we can recall the Stinespring result, \cite{Sti}, 
which is a generalization of GNS construction and which was the starting point for
a general interest in the concept of complete positivity.

\begin{theorem}(\cite{Sti})
{ For an operator state $\tau$ there is a Hilbert space $\cH$, 
a $^*$-representation ($^*$-morphism) $\pi : \cA \to
\cB(\cH)$ and a partial isometry $V : \cK \to \cH$ for which}
\begin{equation}
\label{Stinspring}
\tau(a) = V^* \pi(a) V.
\end{equation}
\end{theorem}

A nice and frequently used criterion for CP can be extracted from Takesaki book \cite{Tak}:

\begin{criterion}
\label{kryt1}
Let $\cA$ and $\cB$ be \Cs-algebras. A linear map $\phi : \cA \to \cB$ is CP if and only if 
\begin{equation}
\sum_{i,j=1}^n y^*_i \phi(x^*_i  x_j)y_j \geq 0
\end{equation}
for every $x_1,...,x_n \in \cA$, $y_1,...,y_n \in \cB$, and every $n \in \bN$.
\end{criterion}

\smallskip

Up to now we considered linear positive maps 
on an algebra without entering into the (possible) complexity of 
the underlying algebra. The situation changes when one is dealing 
with composed systems (for example in the framework of 
open system theory). Namely, 
there is a need to use the tensor product structure.
At this point, it is worth citing Takesaki's remark \cite{Tak}:``...Unlike the finite dimensional case, the tensor product of infinite dimensional Banach spaces behaves mysteriously.''
He had in mind ``topological properties of Banach spaces'' , i.e.: `` cross norms in the tensor product
are highly non-unique.''

But from the point of view of composed systems the situation is, even, more mysterious as finite dimensional cases are also obscure. To explain this point, let us
consider positive maps defined on the tensor product
of two \Cs-algebras, 
$\tau : \cA \otimes \cB \to \cA \otimes \cB$.
But now the question of order is much more complicated. Namely, 
there are various cones determining the order structure in the tensor product of algebras (cf. \cite{Witt})
\begin{equation}
 \C_{inj} \equiv (\cA \otimes \cB)^+ \supseteq, ..., 
\supseteq \C_{\beta} \supseteq,..., \supseteq \C_{pro}\equiv
conv(\cA^+\otimes\cB^+)
\end{equation}
and correspondingly in terms of states (cf \cite{I})
\begin{equation}
\cS(\cA \otimes \cB) \supseteq,..., \supseteq \cS_{\beta}
\supseteq, ..., \supseteq conv(\cS(\cA)\otimes\cS(\cB)).
\end{equation}
Here, $\C_{inj}$ stands for the injective cone, $\C_{\beta}$ for a tensor cone,
while $\C_{pro}$ for the projective cone. The tensor cone $\C_{\beta}$ 
is defined by the property: the canonical bilinear mappings 
$\omega :\cA_h \times \cB_h
\to (\cA_h \otimes \cB_h, \C_{\beta})$ and $\omega^* : \cA^*_h \times \cB^*_h \to 
(\cA^*_h \otimes \cB^*_h, \C_{\beta}^*)$ are positive. 
The cones $\C_{inj}, C_{\beta}, C_{pro}$ are different unless either
$\cA$, or $\cB$, or both $\cA$ and $\cB$ are abelian (so a finite dimension does not help very much!). 
This feature is the origin of various positivity concepts
for non-commutative composed systems and
it was Stinespring who used the partial transposition 
(transposition tensored with identity map) for showing the difference
between $C_{\beta}$ and $\C_{inj}$ and $\C_{pro}$ (see \cite{Witt}, also \cite{Hor}). Clearly, in dual terms, the mentioned property
corresponds to the fact that the set of separable states $conv(\cS(\cA)\otimes\cS(\cB))$ is different from the set of all states
and that there are various special subsets of states if both subsystems are truly quantum.

In his pioneering work on Banach spaces, Grothendieck \cite{Gro} observed the links between tensor products and mapping spaces. A nice example of such links was provided by St{\o}rmer \cite{St1}. To present this result we need a little preparation. 

Let $\mathfrak{A}$ denote a norm closed self-adjoint subspace of bounded operators on a Hilbert space $\mathcal K$ containing the identity operator on $\mathcal K$. $\mathfrak T$ will denote the set of trace class operators on $\cB(\cH)$. $x \to x^t$ denotes the transpose map of $\cB(\cH)$ with respect to some orthonormal basis.
The set of all linear bounded (positive) maps $\phi: \mathfrak{A} \to \cB(\cH)$ will be denoted by $\cB(\mathfrak{A}, \cB(\cH))$ ($\cB(\mathfrak{A}, \cB(\cH))^+$ respectively). 
 Finally, we denote by ${\mathfrak A} \odot {\mathfrak T}$ the algebraic tensor product of $\mathfrak A$ and $\mathfrak T$
 (algebraic tensor product of two vector spaces is defined as its $^*$-algebraic structure when the factor spaces are $^*$-algebras; so the topological questions are not considered) and denote by ${\mathfrak A} \hat{\otimes} \mathfrak T$ its Banach space closure under the projective norm defined by
\begin{equation}
||x|| = \inf \{ \sum_{i=1}^n ||a_i|| ||b_i||_1: x = \sum_{i=1}^n a_i \otimes b_i, \ a_i \in {\mathfrak A}, \ b_i \in {\mathfrak T} \},
\end{equation}
where $|| \cdot ||_1$ stands for the trace norm. Now, we are in a position to give (see \cite{St1})
\begin{lemma}
\label{pierwszy lemat}
There is an isometric isomorphism $\phi \to \tilde{\phi}$ between $\cB({\mathfrak A}, \cB(\cH))$ and
$({\mathfrak A} \hat{\otimes} {\mathfrak T})^*$ given by
\begin{equation}
(\tilde{\phi})(\sum_{i=1}^n a_i\otimes b_i) = \sum_{i=1}^nTr(\phi(a_i)b^t_i),
\end{equation}
where $\sum_{i=1}^n a_i\otimes b_i \in {\mathfrak A}\odot {\mathfrak T}$.

Furthermore, $ \phi \in \cB({\mathfrak A}, \cB(\cH))^+$ if and only if $\tilde{\phi}$ is positive on ${\mathfrak A}^+ \hat{\odot} {\mathfrak T}^+$.
\end{lemma}

To comment this result we make
\begin{remark}
\begin{enumerate}
\item There is not any restriction on the dimension of Hilbert space. In other words, this result can be applied to true quantum systems.
\item In \cite{St2}, St{\o}rmer showed that in the special case when ${\mathfrak A} = M_n(\bC)$ and $\cH$ has dimension equal to $n$, the above Lemma is a reformulation of Choi result \cite{Ch1}, \cite{Ch3}, (see also \cite{Jam}).
\item  One should note that the positivity  of a functional is defined by the projective cone
${\mathfrak A}^+ \hat{\odot} {\mathfrak T}^+$.
\item A generalization of the Choi result (mentioned in 4.2) was also obtained by Belavkin and Staszewski \cite{Slawa}.
\end{enumerate}
\end{remark}

We will also need the concept of co-CP maps. A map $\phi$ is co-CP if its composition with the transposition is a CP map. To see that this is not a trivial condition it is enough to note that the transposition is not even a 2-positive (2-positivity means that the condition given in (\ref{CP}) is satisfied for $n$ equal to $1$ and $2$ only).
A larger class of positive maps is formed by decomposable maps. A map $\phi$ is called decomposable if it can be written as a sum of CP and co-CP maps. Equivalently, if in (\ref{Stinspring}) one replaces $^*$-morphism by Jordan morphism (i.e. a linear map which preserves anticommutator) then the canonical form of a decomposable map is obtained.

Turning to states, it was mentioned that $conv(\cS(\cA)\otimes\cS(\cB))$ are called separable states. The subset of states
$\cS(\cA\otimes \cB) \setminus conv(\cS(\cA)\otimes\cS(\cB))$ is called the set of entangled states. We will be interested in the special subset of states:
\begin{equation}
\cS(\cA \otimes \cB)_{PPT} \equiv \cS_{PPT} = \{ \varphi \in \cS(\cA \otimes \cB); \varphi \circ (t \otimes id) \in \cS(\cA \otimes \cB) \}
\end{equation}
where $t$ stands for transposition. Such states are called PPT states. It is worth observing that the condition in the definition of $\cS_{PPT}$ is non-trivial; namely the partial transposition does not need to be a positive map!
Clearly
\begin{equation*}
\cS \supseteq \cS_{PPT}\supseteq \cS_{sep}.
\end{equation*}

Among entangled states , the states called maximally entangled are of special interest. They can be defined as those for which the state reduced to a subsystem is maximally chaotic (in the entropic sense). A nice example of such states is given by EPR (Einstein-Podolsky-Rosen) states.

\vskip 1cm

The aim of this paper is to give a general characterization of PPT states. We shall present two approaches. The first one is based on the structure of positive maps and as the starting point we will take a modification of Lemma \ref{pierwszy lemat}.
The second approach employs the Hilbert space geometry. This equivalent description will offer rather strikingly very simple definitions of entanglement measures.

\section{Entanglement mappings and PPT states}
In this Section we present a modification of Belavkin-Ohya approach \cite{BO}, \cite{BO2}, \cite{Matsuoka}, and \cite{BD} to the characterization of entanglement.
The basic concept of this approach is the entangling operator $H$. The aim of this section is to provide explicit formulas for both entangling operator $H$ and entanglement mapping $\phi^*$ as well as to give the first characterization of PPT states.

Let us consider a composed system $\sum $ consisting of two subsystems $%
\sum_{1}$, $\sum_{2}$. We assume that  $\sum_{1}$ is defined by the pair $\left( \mathcal{H},%
\mathcal{B}(\mathcal{H})\right) $ while $\sum_{2}$ by the pair $ \left( 
\mathcal{K},\mathcal{B}(\mathcal{K})\right) $ respectively, where $\mathcal{H%
}$ $(\mathcal{K)}$ is a separable Hilbert space. Let $\omega $ be a normal
compound state on $\sum $, i.e. $\omega $ is a normal state on $\mathcal{B}%
\left( \mathcal{H}\otimes \mathcal{K}\right) $. Thus 
\begin{equation*}
\omega \left( a\otimes b\right) =Tr\rho_{\omega} \left( a\otimes b\right)
\end{equation*}%
with $a\in \mathcal{B}\left( \mathcal{H}\right) $, $b\in \mathcal{B}\left( 
\mathcal{K}\right) $. $\rho_{\omega} \equiv \rho$ is a density matrix with the
spectral resolution $\rho =\underset{i}{\sum }\lambda _{i}\left\vert
e_{i}\right\rangle \left\langle e_{i}\right\vert $. Define a linear bounded
operator $T_{\zeta }:\mathcal{K}\rightarrow \mathcal{H}\otimes \mathcal{K}$
by%
\begin{equation}
T_{\zeta }\eta =\zeta \otimes \eta
\end{equation}%
where $\zeta \in \mathcal{H}$, $\eta \in \mathcal{K}$.

Note that the adjoint operator $T_{\zeta }^{\ast }:\mathcal{H}\otimes \mathcal{K}$ $\rightarrow 
\mathcal{K}$ is given by%
\begin{equation}
T_{\zeta }^{\ast }\zeta ^{\prime }\otimes \eta ^{\prime }=\left( \zeta
,\zeta ^{\prime }\right) \eta ^{\prime }.
\end{equation}

Now we wish, following B-O scheme, to define the operator 
\begin{equation*}
H:\mathcal{H}\rightarrow \mathcal{H}\otimes \mathcal{K}\otimes \mathcal{K}
\end{equation*}%
by the formula:
\begin{equation}
H\zeta =\underset{i}{\sum }\lambda _{i}^{\frac{1}{2}}\left( J_{\mathcal{H}%
\otimes \mathcal{K}}\otimes T_{J_{\mathcal{H}}\zeta }^{\ast }\right)
e_{i}\otimes e_{i}
\end{equation}%
where $J_{\mathcal{H}\otimes \mathcal{K}}$ is a complex conjugation defined
by $J_{\mathcal{H}\otimes \mathcal{K}}f\equiv J_{\mathcal{H}\otimes \mathcal{%
K}}\underset{i}({\sum }\left( e_{i}^{\cdot },f\right) e_{i}^{\cdot })=\underset%
{i}{\sum }\overline{\left( e_{i}^{\cdot },f\right) }e_{i}^{\cdot }$ where $%
\left\{ e_{i}^{\cdot }\right\} $ is any CONS extending (if necessary) the
orthogonal system $\left\{ e_{i}\right\} $ determined by the spectral
resolution of $\rho $. ( $J_{\mathcal{H}}$ is defined analogously with the
spectral resolution given by $H^{\ast }H$; using the explicit form of $H$, easy calculations show that the spectrum of $H^*H$ is discrete.) We wish to show

\begin{theorem}
\label{pierwsze}
The normal state $\omega $ can be represented as%
\begin{equation}
\omega \left( a\otimes b\right) =Tr_{\mathcal{H}}a^{t}H^{\ast }\left(
1\otimes b\right) H
\end{equation}%
where a$^{t}=J_{\mathcal{H}}a^{\ast }J_{\mathcal{H}}$.

\begin{proof}
\begin{eqnarray*}
Tra^{t}H^{\ast }\left( 1\otimes b\right) H &=&\underset{k}{\sum }\left(
h_{k},a^{t}H^{\ast }\left( 1\otimes b\right) Hh_{k}\right)  \\
&=&\underset{k}{\sum }\left( h_{k},J_{\mathcal{H}}a^{\ast }J_{\mathcal{H}%
}H^{\ast }\left( 1\otimes b\right) Hh_{k}\right)  \\
&=&\underset{k}{\sum }\left( HJ_{\mathcal{H}}aJ_{\mathcal{H}}h_{k},\left(
1\otimes b\right) Hh_{k}\right)  \\
&=&\underset{k}{\sum }\underset{i,j}{\sum }\left( \lambda _{i}^{\frac{1}{2}%
}\left( J_{\mathcal{H}\otimes \mathcal{K}}\otimes T_{J_{\mathcal{H}}J_{%
\mathcal{H}}aJ_{\mathcal{H}}h_{k}}^{\ast }\right) e_{i}\otimes e_{i},\left(
1\otimes b\right) \lambda _{j}^{\frac{1}{2}}\left( J_{\mathcal{H}\otimes 
\mathcal{K}}\otimes T_{J_{\mathcal{H}}h_{k}}^{\ast }\right) e_{j}\otimes
e_{j}\right)  \\
&=&\underset{k,i,j}{\sum }\lambda _{i}^{\frac{1}{2}}\lambda _{j}^{\frac{1}{2}%
}\left( e_{i}\otimes T_{aJ_{\mathcal{H}}h_{k}}^{\ast }e_{i},\left( 1\otimes
b\right) e_{j}\otimes T_{J_{\mathcal{H}}h_{k}}^{\ast }e_{j}\right). 
\end{eqnarray*}%
Note that $\left\{ h_{k}\right\} $ can be chosen in such the way that it is  CONS used in the definition of $J_{%
\mathcal{H}}$. So%
\begin{eqnarray*}
Tra^{t}H^{\ast }\left( 1\otimes b\right) H &=&\underset{k,i,j}{\sum }\lambda
_{i}^{\frac{1}{2}}\lambda _{j}^{\frac{1}{2}}\left( e_{i}\otimes
T_{ah_{k}}^{\ast }e_{i},\left( 1\otimes b\right) e_{j}\otimes
T_{h_{k}}^{\ast }e_{j}\right)  \\
&=&\underset{k,i}{\sum }\lambda _{i}\left( T_{ah_{k}}^{\ast
}e_{i},bT_{h_{k}}^{\ast }e_{i}\right) .
\end{eqnarray*}%
Let $\left\{ v_{m}\otimes w_{n}\right\} $ be a CONS in $\mathcal{H}\otimes 
\mathcal{K}$. Then%
\begin{eqnarray*}
&&Tra^{t}H^{\ast }\left( 1\otimes b\right) H \\
&=&\underset{k,i,m,n,p,r}{\sum }\lambda _{i}\left( e_{i},v_{m}\otimes
w_{n}\right) \left( T_{ah_{k}}^{\ast }v_{m}\otimes w_{n},bT_{h_{k}}^{\ast
}v_{p}\otimes w_{r}\right) \left( v_{p}\otimes w_{r},e_{i}\right)  \\
&=&\underset{k,i,m,n,p,r}{\sum }\lambda _{i}\left( e_{i},v_{m}\otimes
w_{n}\right) \left( v_{p}\otimes w_{r},e_{i}\right) \overline{\left(
ah_{k},v_{m}\right) }\left( h_{k},v_{p}\right) \left( w_{n},bw_{r}\right). 
\end{eqnarray*}%
As $\left\{ v_{m}\right\} $ is a CONS in $\mathcal{H}$ we can take $\left\{
v_{m}\right\} =\left\{ h_{m}\right\} $. So that%
\begin{eqnarray*}
&&Tra^{t}H^{\ast }\left( 1\otimes b\right) H \\
&=&\underset{k,i,m,n,p,r}{\sum }\lambda _{i}\left( e_{i},h_{m}\otimes
w_{n}\right) \left( h_{p}\otimes w_{r},e_{i}\right) \overline{\left(
ah_{k},h_{m}\right) }\left( h_{k},h_{p}\right) \left( w_{n},bw_{r}\right)  \\
&=&\underset{k,i,m,n,r}{\sum }\lambda _{i}\left( e_{i},h_{m}\otimes
w_{n}\right) \left( h_{k}\otimes w_{r},e_{i}\right) \left(
h_{m},ah_{k}\right) \left( w_{n},bw_{r}\right)  \\
&=&\underset{k,i,m,n,r}{\sum }\lambda _{i}\left( e_{i},h_{m}\otimes
w_{n}\right) \left( h_{m}\otimes w_{n},\left( a\otimes b\right) h_{k}\otimes
w_{r}\right) \left( h_{k}\otimes w_{r},e_{i}\right)  \\
&=&\underset{i}{\sum }\lambda _{i}\left( e_{i},\left( a\otimes b\right)
e_{i}\right) =Tr\rho \left( a\otimes b\right) =\omega \left( a\otimes
b\right) .
\end{eqnarray*}
\end{proof}
\end{theorem}

As it was mentioned in the previous Section,
in his fundamental paper on topological linear spaces, Grothendieck
emphasized the importance of relating mapping space to tensor products. Another nice 
example of such relations is given by the following modification of St\o rmer's result.
\begin{lemma}
\label{drugi lemat}
(1) Let
$\mathcal{B}\left[ \ \mathcal{B}\left( \mathcal{H}\right) ,\mathcal{B(K)}%
_{\ast }\right] $ stand for the set of all linear, bounded, normal (so weakly$^*$-continuous) maps from $\cB(\cH)$ into $\cB(\cH)_{\ast}$.
There is an isomorphism $\psi \longmapsto \Psi $ between $%
\mathcal{B}\left[ \ \mathcal{B}\left( \mathcal{H}\right) ,\mathcal{B(K)}%
_{\ast }\right] $ and $\left( \mathcal{B(H)\otimes B(K)}\right)_{\ast }$
given by 
\begin{equation}
\Psi \left( \sum_{i}a_{i}\otimes b_{i}\right) =\sum_{i}Tr_{\mathcal{K}}\psi
\left( a_{i}\right) b_{i}^{t},\text{\ \ }a_{i}\in \mathcal{B}\left( \mathcal{%
H}\right) ,\text{ }b_{i}\in \mathcal{B}\left( \mathcal{K}\right) .
\end{equation}%
The isomorphism is isometric if $\Psi$ is considered on $\mathcal{B(H)\hat{\otimes} B(K)}$. 
Furthermore $\Psi $ is positive on $(\cB(\cH)\otimes \cB(\cK))^+$ iff $\psi $ is complete positive.

(2) There is an isomorphism $\phi \longmapsto \Phi $ between $%
\mathcal{B}\left[ \ \mathcal{B}\left( \mathcal{H}\right) ,\mathcal{B(K)}%
_{\ast }\right] $ and $\left( \mathcal{B(H)\otimes B(K)}\right)_{\ast }$
given by 
\begin{equation}
\Phi \left( \sum_{i}a_{i}\otimes b_{i}\right) =\sum_{i}Tr_{\mathcal{K}}\phi
\left( a_{i}\right) b_{i},\text{\ \ }a_{i}\in \mathcal{B}\left( \mathcal{H}%
\right) ,\text{ }b_{i}\in \mathcal{B}\left( \mathcal{K}\right) .
\end{equation}%
The isomorphism is isometric if $\Phi$ is considered on $\mathcal{B(H)\hat{\otimes} B(K)}$. 
Furthermore $\Phi $ is positive on $(\cB(\cH)\otimes \cB(\cK))^+$ iff $\phi $ is complete co-positive.
\end{lemma}
\begin{proof}
A repetition of modified St{\o}rmer's and standard arguments (cf \ref{7}.2-3 below).
\end{proof}

We want to comment this lemma with
\begin{remark}
\label{7}
\begin{enumerate}
\item Firstly, one should note the  basic difference between Lemma \ref{pierwszy lemat} and Lemma {\ref{drugi lemat}}.
In Lemma \ref{pierwszy lemat}, the order is defined by the projective cone while in Lemma \ref{drugi lemat}, the order is defined by the injective cone.
\item Secondly, as $\mathcal{B(K)_*}$ is isomorphic to the set of all trace class operators $\mathfrak{T}\equiv
\mathfrak{T}_{\mathcal{K}}$ on $\cK$, $\mathcal{B(B(H),B(K)_*)}$ can be considered as $\cB(\cB(\cH), \mathfrak{T})$.
\item Thirdly, let $\cA$ and $\cB$ be $C^*$-algebras. Lemma \ref{pierwszy lemat} and Lemma \ref{drugi lemat} stem from the standard identification $\Psi \to \psi$ of $(\cA\odot \cB)^d$ with the set $Hom(\cA, \cB^d)$ of linear maps from $\cA$ to $\cB^d$ where $[\psi(a)](b) = \Psi(a\otimes b)$. Here $(\cA \odot \cB)^d$ ($\cB^d$) stands for the algebraic dual of $\cA\odot \cB$ ($\cB$ respectively), see \cite{Bla} for details.
\item Finally, Lemma \ref{drugi lemat} gains in interest if we realize that the operator $H$ defined in
the first part of this section can be used for a definition of the entanglement mapping $\phi
:\mathcal{B}\left( \mathcal{K}\right) \rightarrow \mathcal{B}\left( \mathcal{%
H}\right) _{\ast }$.
\end{enumerate}
\end{remark}

Let us define%
\begin{equation}
\label{la}
\phi \left( b\right) =\left( H^{\ast }\left( 1\otimes b\right) H\right)
^{t}=J_{\mathcal{H}}H^{\ast }\left( 1\otimes b\right) ^{\ast }HJ_{\mathcal{H}.%
}
\end{equation}
Then

\begin{proposition}
\label{PPT}
The entanglement mapping

(i) $\phi ^{\ast }:\mathcal{B}\left( \mathcal{H}\right) \rightarrow \mathcal{%
B}\left( \mathcal{K}\right) _{\ast }$ has the following explicit form%
\begin{equation}
\phi ^{\ast }\left( a\right) =Tr_{\mathcal{H}\otimes \mathcal{K}%
}Ha^{t}H^{\ast }
\end{equation}

(ii) The state $\omega $ on $\mathcal{B}\left( \mathcal{H}\otimes \mathcal{K}%
\right) $ can be written as 
\begin{equation}
\omega \left( a\otimes b\right) =Tr_{\mathcal{H}}a\phi \left( b\right) =Tr_{%
\mathcal{K}}b\phi ^{\ast }\left( a\right)
\end{equation}
where $\phi$ was defined in (\ref{la}).

\begin{proof}
For $f$, $g\in \mathcal{K}$ and $h\in \mathcal{H}$%
\begin{eqnarray*}
Tr_{\mathcal{K}}\phi ^{\ast }\left( a\right) \left\vert f\right\rangle
\left\langle g\right\vert  &=&\left( g,\phi ^{\ast }\left( a\right) f\right) 
\\
&=&\underset{i}{\sum }\left( e_{i}\otimes g,Ha^{t}H^{\ast }e_{i}\otimes
f\right), 
\end{eqnarray*}%
where as before $\{ e_i \}$ is a CONS in $\cH \otimes \cK$.
Note:%
\begin{eqnarray*}
\left( h,H^{\ast }e_{i}\otimes f\right)  &=&\left( Hh,e_{i}\otimes f\right) 
\\
&=&\underset{k}{\sum }\left( \lambda _{k}^{\frac{1}{2}}\left( J_{\mathcal{H}%
\otimes \mathcal{K}}\otimes T_{J_{\mathcal{H}}h}^{\ast }\right) e_{k}\otimes
e_{k},e_{i}\otimes f\right)  \\
&=&\underset{k}{\sum }\lambda _{k}^{\frac{1}{2}}\left( e_{k}\otimes T_{J_{%
\mathcal{H}}h}^{\ast }e_{k},e_{i}\otimes f\right)  \\
&=&\underset{k,m,n}{\sum }\lambda _{k}^{\frac{1}{2}}\left(
e_{k},v_{m}\otimes w_{n}\right) \left( e_{k}\otimes \left( J_{\mathcal{H}%
}h,v_{m}\right) w_{n},e_{i}\otimes f\right) 
\end{eqnarray*}%
where $\left\{ v_{m}\right\} $ is a CONS in $\mathcal{H}$ such that $J_{%
\mathcal{H}}$ is defined w.r.t this basis, and $\left\{ w_{n}\right\} $ is a
CONS in $\mathcal{K},$%
\begin{eqnarray*}
&=&\underset{m,n}{\sum }\lambda _{i}^{\frac{1}{2}}\left( e_{i},v_{m}\otimes
w_{n}\right) \overline{\left( J_{\mathcal{H}}h,v_{m}\right) }\left(
w_{n},f\right)  \\
&=&\underset{m,n}{\sum }\lambda _{i}^{\frac{1}{2}}\left( e_{i},v_{m}\otimes
w_{n}\right) \left( v_{m},J_{\mathcal{H}}h\right) \left( w_{n},f\right)  \\
&=&\underset{m,n}{\sum }\lambda _{i}^{\frac{1}{2}}\left( e_{i},v_{m}\otimes
w_{n}\right) \left( v_{n}\otimes w_{n},J_{\mathcal{H}}h\otimes f\right)  \\
&=&\lambda _{i}^{\frac{1}{2}}\left( e_{i},J_{\mathcal{H}}h\otimes f\right). 
\end{eqnarray*}%
In particular, putting $h^{\prime}=\left( a^{t}\right) ^{\ast }H^{\ast }e_{i}\otimes g$ one has%
\begin{eqnarray*}
\left( h^{\prime},v_{m}\right)  &=&\left( H^{\ast }e_{i}\otimes g,a^{t}v_{m}\right) 
\\
&=&\lambda _{i}^{\frac{1}{2}}\left( J_{\mathcal{H}}a^{t}v_{m}\otimes
g,e_{i}\right). 
\end{eqnarray*}%
Hence 
\begin{eqnarray*}
Tr_{\mathcal{K}}\phi ^{\ast }\left( a\right) \left\vert f\right\rangle
\left\langle g\right\vert  &=&\underset{i}{\sum }\left( (a^{t})^*H^{\ast
}e_{i}\otimes g,H^{\ast }e_{i}\otimes f\right)  \\
&=&\underset{i,m,n}{\sum }((a^{t})^*H^{\ast }e_{i}\otimes g,v_{m})(v_{m},H^{\ast
}e_{i}\otimes w_{n})(w_{n},f) \\
&=&\underset{i}{\sum }\lambda _{i}^{\frac{1}{2}}\left( J_{\mathcal{H}%
}a^{t}v_{m}\otimes g,e_{i}\right) \lambda _{i}^{\frac{1}{2}}\left(
e_{i},v_{m}\otimes w_{n}\right) (w_{n},f) \\
&=&\underset{i,m}{\sum }\lambda _{i}\left( J_{\mathcal{H}}a^{t}v_{m}\otimes
g,e_{i}\right) \left( e_{i},v_{m}\otimes f\right)  \\
&=&Tr\rho _{\omega }\left( \underset{m}{\sum }\left\vert v_{m}\otimes
f\right\rangle \left\langle J_{\mathcal{H}}a^{t}v_{m}\otimes g\right\vert
\right)  \\
&=&Tr\rho _{\omega }\left( \underset{m}{\sum }\left\vert v_{m}\otimes
f\right\rangle \left\langle a^*v_{m}\otimes g\right\vert \right)  \\
&=&Tr\rho _{\omega }\left( a\otimes \left\vert f\right\rangle \left\langle
g\right\vert \right) =\omega \left( a\otimes \left\vert f\right\rangle
\left\langle g\right\vert \right) 
\end{eqnarray*}%
Thus 
\begin{equation*}
Tr_{\mathcal{K}}b\phi ^{\ast }\left( a\right) =\omega \left( a\otimes
b\right). 
\end{equation*}

The rest follows from Theorem \ref{pierwsze}.
\end{proof}
\end{proposition}

Theorem \ref{pierwsze}, Lemma \ref{drugi lemat} and Proposition \ref{PPT} lead to

\begin{corollary}
PPT states are completely characterized by entanglement mappings $\phi^{\ast}$ which are both CP and co-CP.
\end{corollary}

This conclusion can be rephrased in the following way (cf \cite{Matsuoka}):
Entanglement mapping $\phi^{\ast}$ which is not CP will be called $q$-entanglement. The set of all $q$-entanglements will be denoted by ${\mathcal E}_q$.
Then PPT criterion can be formulated as:

\begin{corollary}
A state is PPT if and only if its associated entanglement mapping $\phi^{\ast}$ is not in ${\mathcal E}_q$.
\end{corollary}
\section{Examples}

To illustrate the strategy of B-O entanglement maps as well as to get better understanding of positive maps
we present some examples.

\textbf{Example 1}: Let $\omega :\mathcal{B}\left( \mathcal{H}\otimes \mathcal{K}%
\right) \rightarrow \mathbb{C}$ be \textbf{a pure product state}, i.e. 
\begin{eqnarray*}
\omega \left( a\otimes b\right)  &=&\omega _{x\otimes y}\left( a\otimes
b\right)  \\
&\equiv &\left( x\otimes y,\left( a\otimes b\right) x\otimes y\right)  \\
\text{ \ \ \ \ \ \ \ \ \ \ \ \ \ \ \ \ } &=&\left( x,ax\right) (y,by)
\end{eqnarray*}%
where $x\in \cH,y\in \cK$ and $\left\Vert x\right\Vert =1=\left\Vert
y\right\Vert $. Then%
\begin{eqnarray*}
H\zeta  &=&J_{\mathcal{H}\otimes \mathcal{K}}\otimes T_{J_{\mathcal{H}}\zeta
}^{\ast }\left( x\otimes y\right) \otimes \left( x\otimes y\right)  \\
&=&J_{\mathcal{H}\otimes \mathcal{K}}\left( x\otimes y\right) \otimes \left(
J_{\mathcal{H}}\zeta ,x\right) y.
\end{eqnarray*}%
For $f\in \mathcal{H}\otimes \mathcal{K}$, $g\in \mathcal{K}$, $h\in 
\mathcal{H}$ we have%
\begin{eqnarray*}
\left( h,H^{\ast }f\otimes g\right)  &=&\left( Hh,f\otimes g\right)  \\
&=&\left( J_{\mathcal{H}\otimes \mathcal{K}}\left( x\otimes y\right) \otimes
\left( J_{\mathcal{H}}h,x\right) y,f\otimes g\right)  \\
&=&\left( J_{\mathcal{H}\otimes \mathcal{K}}\left( x\otimes y\right)
,f\right) \left( h,J_{\mathcal{H}}x\right) \left( y,g\right)  \\
&=&\left( h,\left( y,g\right) \left( J_{\mathcal{H}\otimes \mathcal{K}%
}\left( x\otimes y\right) ,f\right) J_{\mathcal{H}}x\right)  \\
&=&\left( h,\left( J_{\mathcal{H}\otimes \mathcal{K}}\left( x\otimes
y\right) \otimes y,f\otimes g\right) J_{\mathcal{H}}x\right) .
\end{eqnarray*}%
so that%
\begin{equation*}
H^{\ast }f\otimes g=\left( J_{\mathcal{H}\otimes \mathcal{K}}\left( x\otimes
y\right) \otimes y,f\otimes g\right) J_{\mathcal{H}}x.
\end{equation*}%
Let $v$, $z\in \mathcal{K}$ and $\left\{ e_{i}\right\} $ is a CONS in $%
\mathcal{H}\otimes \mathcal{K}$ then, using the above calculation, we have%
\begin{eqnarray*}
&&\left( v,\phi ^{\ast }\left( a\right) z\right)  \\
&=&\left( v,Tr_{\cH\otimes \cK}Ha^{t}H^{\ast }z\right)  \\
&=&\underset{i}{\sum }\left( e_{i}\otimes v,Ha^{t}H^{\ast }e_{i}\otimes
z\right)  \\
&=&\underset{i}{\sum }\left( H^{\ast }e_{i}\otimes v,a^{t}H^{\ast
}e_{i}\otimes z\right)  \\
&=&\underset{i}{\sum }\left( \left( J_{\mathcal{H}\otimes \mathcal{K}}\left(
x\otimes y\right) \otimes y,e_{i}\otimes v\right) J_{\mathcal{H}%
}x,a^{t}\left( J_{\mathcal{H}\otimes \mathcal{K}}\left( x\otimes y\right)
\otimes y,e_{i}\otimes z\right) J_{\mathcal{H}}x\right)  \\
&=&\underset{i}{\sum }\left( J_{\mathcal{H}\otimes \mathcal{K}}\left(
x\otimes y\right) \otimes y,e_{i}\otimes z\right) \left( e_{i}\otimes v,J_{%
\mathcal{H}\otimes \mathcal{K}}\left( x\otimes y\right) \otimes y\right)
\left( J_{\mathcal{H}}x,a^{t}J_{\mathcal{H}}x\right).
\end{eqnarray*}%
Note that 
\begin{eqnarray*}
\left( J_{\mathcal{H}}x,a^{t}J_{\mathcal{H}}x\right)  &=&\left( J_{\mathcal{H%
}}x,J_{\mathcal{H}}a^{\ast }x\right)  \\
&=&\left( a^{\ast }x,J_{\mathcal{H}}J_{\mathcal{H}}x\right)  \\
&=&\left( x,ax\right). 
\end{eqnarray*}%
Thus%
\begin{eqnarray*}
&&\left( v,\phi ^{\ast }\left( a\right) z\right)  \\
&=&\left( J_{\mathcal{H}\otimes \mathcal{K}}\left( x\otimes y\right) ,J_{%
\mathcal{H}\otimes \mathcal{K}}\left( x\otimes y\right) \right) \left(
y,z\right) \left( v,y\right) \left( x,ax\right)  \\
&=&\left( v,\left\Vert J_{\mathcal{H}\otimes \mathcal{K}}\left( x\otimes
y\right) \right\Vert ^{2}\left( y,z\right) \left( x,ax\right) y\right), 
\end{eqnarray*}%
so that%
\begin{eqnarray*}
\phi ^{\ast }\left( a\right) z &=&\left\Vert x\otimes y\right\Vert
^{2}\left( y,z\right) \left( x,ax\right) y \\
&=&\left( x,ax\right) \left\vert y\right\rangle \left\langle y\right\vert
\cdot z.\text{ \ (because of }\left\Vert x\right\Vert =1=\left\Vert
y\right\Vert .)
\end{eqnarray*}%
Put $P_{y}=\left\vert y\right\rangle \left\langle y\right\vert $. Then, we have%
\begin{equation}
\phi ^{\ast }\left( a\right) =\left( x,ax\right) P_{y}.
\end{equation}

To analyse CP and co-CP property let us 
observe that (we are applying Criterion \ref{kryt1}): $\forall $ $w\in \mathcal{K}$%
\begin{eqnarray*}
\left( w,\underset{i,j}{\sum }b_{i}^{\ast }\phi ^{\ast }\left( a_{i}^{\ast
}a_{j}\right) b_{j}w\right) &=&\underset{i,j}{\sum }\left( x,a_{i}^{\ast
}a_{j}x\right) \left( w,b_{i}^{\ast }y\right) \left( b_{j}^{\ast }y,w\right)
\\
&=&\left( \underset{i}{\sum }\overline{\lambda _{i}}a_{i}x,\underset{j}{\sum 
}\overline{\lambda _{j}}a_{j}x\right) \geq 0,
\end{eqnarray*}%
where $\lambda _{i}=\left( w,b_{i}^{\ast }y\right) .$ Also%
\begin{eqnarray*}
\left( w,\underset{i,j}{\sum }b_{i}^{\ast }\phi ^{\ast }\left( a_{j}^{\ast
}a_{i}\right) b_{j}w\right) &=&\underset{i,j}{\sum }\left( x,a_{j}^{\ast
}a_{i}x\right) \left( w,b_{i}^{\ast }y\right) \left( b_{j}^{\ast }y,w\right)
\\
&=&\left( \underset{j}{\sum }\lambda _{j}a_{j}x,\underset{i}{\sum }\lambda
_{i}a_{i}x\right) \geq 0.
\end{eqnarray*}%
So $\phi ^{\ast }$ is both CP and co-CP. This was expected because any pure separable state is a PPT state.

\vskip 1cm

\textbf{Example 2}:  \textbf{Separable states}: Let $\omega =\underset{i}{\sum }\lambda _{i}\omega
_{x_{i}\otimes y_{i}}$. One has%
\begin{eqnarray*}
\omega \left( a\otimes b\right) &=&\underset{i}{\sum }\lambda _{i}\omega
_{x_{i}\otimes y_{i}}\left( a\otimes b\right) \\
&=&\underset{i}{\sum }\lambda _{i}Tr_{\mathcal{K}}b\phi _{i}^{\ast }\left(
a\right) \\
&=&Tr_{\mathcal{K}}b\underset{i}{\sum }\lambda _{i}\phi _{i}^{\ast }\left(
a\right) \\
&=&Tr_{\mathcal{K}}b\phi ^{\ast }\left( a\right)
\end{eqnarray*}%
where $\phi ^{\ast }=\underset{i}{\sum }\lambda _{i}\phi _{i}^{\ast }$. But $%
\phi _{i}^{\ast }$ was described in Example 1 and is both CP and co-CP so
 $\phi ^{\ast }$ also has this property. Clearly, the conclusion given at the end of Example 1 is also valid here. 

\vskip 1cm

\textbf{Example 3}: \textbf{A pure state}. Let $\omega $ be a pure state on $%
\mathcal{B}\left( \mathcal{H}\otimes \mathcal{K}\right) $. As any pure state
on the factor I is a vector state so there exists $x\in \mathcal{H}\otimes 
\mathcal{K}$ such that 
\begin{equation*}
\omega \left( a\otimes b\right) =\left( x,\left( a\otimes b\right) x\right) .
\end{equation*}%
Let $z\in \mathcal{H}\otimes \mathcal{K}$, $h$, $\zeta \in \mathcal{H}$, $%
g\in \mathcal{K}$ and $\left\{ v_{i}\right\} $ be CONS in $\mathcal{H}$, $%
\left\{ w_{j}\right\} $ be CONS in $\mathcal{K}$. Then 
\begin{equation*}
H\zeta =\left( J_{\mathcal{H}\otimes \mathcal{K}}\otimes T_{J_{\mathcal{H}%
}\zeta }^{\ast }\right) \left( x\otimes x\right) =J_{\mathcal{H}\otimes 
\mathcal{K}}x\otimes T_{J_{\mathcal{H}}\zeta }^{\ast }x.
\end{equation*}%
Also%
\begin{equation*}
\left( h,H^{\ast }z\otimes y\right) =\left( Hh,z\otimes y\right) =\left( J_{%
\mathcal{H}\otimes \mathcal{K}}x,z\right) \left( T_{J_{\mathcal{H}}h}^{\ast
}x,y\right)
\end{equation*}%
where 
\begin{eqnarray*}
T_{J_{\mathcal{H}}h}^{\ast }x &=&\underset{i,j}{\sum }T_{J_{\mathcal{H}%
}h}^{\ast }\left( v_{i}\otimes w_{j},x\right) v_{i}\otimes w_{j} \\
&=&\underset{i,j}{\sum }\left( v_{i}\otimes w_{j},x\right) \left( J_{%
\mathcal{H}}h,v_{i}\right) w_{j}.
\end{eqnarray*}%
Thus%
\begin{eqnarray}
\label{5}
\left( h,H^{\ast }z\otimes y\right) &=&\underset{i,j}{\sum }\left( J_{%
\mathcal{H}\otimes \mathcal{K}}x,z\right) \left( x,v_{i}\otimes w_{j}\right)
\left( v_{i},J_{\mathcal{H}}h\right) \left( w_{j},y\right)  \notag \\
&=&\underset{i,j}{\sum }\left( J_{\mathcal{H}\otimes \mathcal{K}}x,z\right)
\left( x,v_{i}\otimes w_{j}\right) \left( v_{i}\otimes w_{j},J_{\mathcal{H}%
}h\otimes y\right)  \notag \\
&=&\left( J_{\mathcal{H}\otimes \mathcal{K}}x,z\right) \overline{\left( J_{%
\mathcal{H}}h\otimes y,x\right) }.
\end{eqnarray}%
Let $\left\{ e_{i}\right\} $ be CONS in $\mathcal{H}\otimes \mathcal{K}$
then, for $w$, $u\in \mathcal{K}$%
\begin{eqnarray*}
\left( w,\phi ^{\ast }\left( a\right) u\right) &=&\left( w,\left( Tr_{%
\mathcal{H\otimes K}}Ha^{t}H^{\ast }\right) u\right) \\
&=&\underset{n}{\sum }\left( e_{n}\otimes w,Ha^{t}H^{\ast }e_{n}\otimes
u\right) \\
&=&\underset{n}{\sum }\left( H^{\ast }e_{n}\otimes w,a^{t}H^{\ast
}e_{n}\otimes u\right).
\end{eqnarray*}%
Let us use (\ref{5}), i.e. put $h=a^{t}H^{\ast }e_{n}\otimes u$, $z=e_{n}$, $y=w$. Then one gets%
\begin{eqnarray*}
&&\left( w,\phi ^{\ast }\left( a\right) u\right) \\
&=&\underset{n}{\sum }\left( e_{n},J_{\mathcal{H}\otimes \mathcal{K}%
}x\right) \left( J_{\mathcal{H}}\left( a^{t}H^{\ast }e_{n}\otimes u\right)
\otimes w,\underset{i,j}{\sum }\left( v_{i}\otimes w_{j},x\right)
v_{i}\otimes w_{j}\right) \\
&=&\underset{n,i,j}{\sum }\left( e_{n},J_{\mathcal{H}\otimes \mathcal{K}%
}x\right) \left( v_{i}\otimes w_{j},x\right) \left( J_{\mathcal{H}}\left(
a^{t}H^{\ast }e_{n}\otimes u\right) ,v_{i}\right) \left( w,w_{j}\right) \\
&=&\underset{n,i,j}{\sum }\left( e_{n},J_{\mathcal{H}\otimes \mathcal{K}%
}x\right) \overline{\left( x,v_{i}\otimes w_{j}\right) }\overline{\left(
w_{j},w\right) }\overline{\left( \left( a^{t}H^{\ast }e_{n}\otimes u\right)
,J_{\mathcal{H}}v_{i}\right) } \\
&=&\underset{n,i}{\sum }\left( e_{n},J_{\mathcal{H}\otimes \mathcal{K}%
}x\right) \left( v_{i}\otimes w,x\right) \left( \left( a^{t}\right) ^{\ast
}J_{\mathcal{H}}v_{i},H^{\ast }e_{n}\otimes u\right)
\end{eqnarray*}%
Again, using (\ref{5}), i.e. putting $h=\left( a^{t}\right) ^{\ast }J_{\mathcal{H}}v_{i}$, $%
z=e_{n}$, $y=u$, one gets

\begin{eqnarray*}
&&\left( w,\phi ^{\ast }\left( a\right) u\right) \\
&=&\underset{n,i}{\sum }\left( e_{n},J_{\mathcal{H}\otimes \mathcal{K}%
}x\right) \left( v_{i}\otimes w,x\right) \left( J_{\mathcal{H}\otimes 
\mathcal{K}}x,e_{n}\right) \overline{\left( J_{\mathcal{H}}\left(
a^{t}\right) ^{\ast }J_{\mathcal{H}}v_{i}\otimes u,x\right) } \\
&=&\underset{n,i}{\sum }\left( J_{\mathcal{H}\otimes \mathcal{K}%
}x,e_{n}\right) \left( e_{n},J_{\mathcal{H}\otimes \mathcal{K}}x\right)
\left( v_{i}\otimes w,x\right) \left( x,av_{i}\otimes u\right) \text{ \ }(%
\text{because of }\left( a^{t}\right) ^{t}=a) \\
&=&\underset{i}{\sum }\left( x,av_{i}\otimes u\right) \left( v_{i}\otimes
w,x\right) \text{\ \ \ \ \ \ (as }\omega _{x}\text{ is a state, }\left\Vert
x\right\Vert =1) \\
&=&\left( x,\left( a\otimes 1\right) \underset{i}{\sum }\left\vert
v_{i}\right\rangle \left\langle v_{i}\right\vert \otimes \left\vert
u\right\rangle \left\langle w\right\vert x\right) \\
&=&\left( x,\left( a\otimes \left\vert u\right\rangle \left\langle
w\right\vert \right) x\right).
\end{eqnarray*}%
Consequently 
\begin{equation}
Tr_{\mathcal{K}}\phi ^{\ast }\left( a\right) \left\vert u\right\rangle
\left\langle w\right\vert =Tr_{\mathcal{H}\otimes \mathcal{K}}\left(
a\otimes \left\vert u\right\rangle \left\langle w\right\vert \right) P_{x}.
\end{equation}%
This means that:
\begin{equation}
\label{stanczysty}
\phi ^{\ast }\left( a\right) =Tr_{\mathcal{H}}\left( a\otimes 1\right) P_{x}.
\end{equation}

Turning to the analysis of CP and co-CP we begin with co-CP property. 
To this end let $\left\{ v_{k}\right\} $ be CONS in $
\mathcal{H}$ then 
\begin{eqnarray*}
\left( w,\underset{i,j}{\sum }b_{i}^{\ast }\phi ^{\ast }\left( a_{j}^{\ast
}a_{i}\right) b_{j}w\right) &=&\underset{i,j}{\sum }\left( x,\left(
a_{j}^{\ast }a_{i}\otimes \left\vert b_{j}w\right\rangle \left\langle
b_{i}w\right\vert \right) x\right) \\
&=&\underset{i,j}{\sum }\left( x,\left( a_{j}^{\ast }\left( \underset{k}{%
\sum }\left\vert v_{k}\right\rangle \left\langle v_{k}\right\vert \right)
a_{i}\otimes \left\vert b_{j}w\right\rangle \left\langle b_{i}w\right\vert
\right) x\right) \\
&=&\underset{i,j,k}{\sum }\left( x,a_{j}^{\ast }v_{k}\otimes b_{j}w\right)
\left( a_{i}^{\ast }v_{k}\otimes b_{i}w,x\right) \\
&=&\underset{k}{\sum }\left( x,\underset{j}{\sum }a_{j}^{\ast }v_{k}\otimes
b_{j}w\right) \left( \underset{i}{\sum }a_{i}^{\ast }v_{k}\otimes
b_{i}w,x\right) \geq 0.
\end{eqnarray*}%
Thus $\phi ^{\ast }$ is a co-CP map.

Now, consider CP condition: Let $\left\{ v_{k}\right\} $ be CONS in $%
\mathcal{H}$ and $\left\{ z_{l}\right\} $ be CONS in $\mathcal{K}$. Assume that $x$ is given by 
\begin{equation*}
x=\underset{k}{\sum }\lambda _{k}v_{k}\otimes z_{k}\text{\ \ }\left( \lambda
_{k}\in \mathbb{C},\underset{k}{\sum }\left\vert \lambda _{k}\right\vert
^{2}=1\right)
\end{equation*}%
where at least two elements of $\left\{ \lambda _{k}\right\} $ are non-zero.
In order to show the non-CP of $\phi ^{\ast }$ some preliminaries are necessary. We recall that $M_n(\cA)$
denotes the \Cs-algebra of $ n \times n$ matrices with entries in $\cA$. Let $\{ e_{ij}\}$ be the canonical basis for $M_n(\bC)\equiv M_n $, i.e. the $ n \times n$ matrices with a $``1''$ in row $i$, column $j$, and zeros elsewhere. It is well known that every element $y$ in $\cA \odot M_n$ can be written
\begin{equation}
\label{22}
y = \sum a_{ij} \otimes e_{ij}
\end{equation}
where the $a_{ij}$'s (being in $\cA$) are unique. The map
\begin{equation}
\label{23}
\Theta: \cA \odot M_n \to M_n(\cA): \sum a_{ij} \otimes e_{ij} \mapsto \{a_{ij} \}
\end{equation}
is linear, multiplicative, $^*$-preserving, and bijective. Therefore, it should be clear that
the complete positivity of $\phi ^{\ast }$ is equivalent to the
positivity of operator   $\sum_{i,j=1}^n e_{ij} \otimes \phi ^{\ast }(a^*_i a_j)$, for any $n$.

 Let $\left\{
e_{i}\right\} $ be CONS in $\mathbb{C}^{n}$. One has%
\begin{eqnarray*}
\phi ^{\ast }\left( a_{i}^{\ast }a_{j}\right) &=&Tr_{\mathcal{H}}\left(
a_{i}^{\ast }a_{j}\otimes 1\right) P_{x} \\
&=&\underset{k.l}{\sum }Tr_{\mathcal{H}}\left( a_{i}^{\ast }a_{j}\otimes
1\right) \left\vert \lambda _{k}v_{k}\otimes z_{k}\right\rangle \left\langle
\lambda _{l}v_{l}\otimes z_{l}\right\vert \\
&=&\underset{k.l}{\sum }\lambda _{k}\overline{\lambda _{l}}Tr_{\mathcal{H}%
}\left( a_{i}^{\ast }a_{j}\left\vert v_{k}\right\rangle \left\langle
v_{l}\right\vert \right) \left\vert z_{k}\right\rangle \left\langle
z_{l}\right\vert \\
&=&\underset{k.l}{\sum }\lambda _{k}\overline{\lambda _{l}}\left(
v_{l},a_{i}^{\ast }a_{j}v_{k}\right) \left\vert z_{k}\right\rangle
\left\langle z_{l}\right\vert
\end{eqnarray*}%
Thus 
\begin{equation*}
\underset{i,j=1}{\overset{n}{\sum }}\left\vert e_{i}\right\rangle
\left\langle e_{j}\right\vert \otimes \phi ^{\ast }\left( a_{i}^{\ast
}a_{j}\right) 
= \underset{i,j=1}{\overset{n}{\sum }}\underset{k.l}{\sum }\lambda _{k}%
\overline{\lambda _{l}}\left( v_{l},a_{i}^{\ast }a_{j}v_{k}\right)
\left\vert e_{i}\right\rangle \left\langle e_{j}\right\vert \otimes
\left\vert z_{k}\right\rangle \left\langle z_{l}\right\vert .
\end{equation*}%
Put $a_{i}=\left\vert y\right\rangle $ $\left\langle v_{i}\right\vert $ \ ( $%
y\in \mathcal{H}$, $\left\Vert y\right\Vert =1$) then%
\begin{equation*}
\{ \phi ^{\ast }\left( \left\vert v_{i}\right\rangle \left\langle
v_{j}\right\vert \right) \} \cong \underset{i,j=1}{\overset{n}{\sum }}%
\lambda _{j}\overline{\lambda _{i}}\left\vert e_{i}\right\rangle
\left\langle e_{j}\right\vert \otimes \left\vert z_{j}\right\rangle
\left\langle z_{i}\right\vert
\end{equation*}%
The positivity of $\{ \phi ^{\ast }\left( \left\vert v_{i}\right\rangle
\left\langle v_{j}\right\vert \right) \}$ means the positivity of $%
\left( \Psi ,(\sum \phi ^{\ast }\left( \left\vert v_{i}\right\rangle
\left\langle v_{j}\right\vert \right) \otimes e_{ij})\Psi \right) $ for any $%
\Psi \in \mathbb{C}^{n}\otimes \mathcal{K}$. Let us take $\Psi _{\pm }$ in the form
\begin{equation*}
\Psi _{\pm }=e_{k}\otimes z_{l}\pm e_{l}\otimes z_{k}.
\end{equation*}%
and assume that $k\neq l$.
Then%
\begin{equation*}
\left( \Psi _{\pm },(\sum \phi ^{\ast }\left( \left\vert v_{i}\right\rangle
\left\langle v_{j}\right\vert \right) \otimes e_{ij})\Psi _{\pm }\right) =\pm 2%
{Re}\lambda _{k}\overline{\lambda _{l}}\in \mathbb{R}\text{.}
\end{equation*}%
If $2Re\lambda _{k}\overline{\lambda _{l}}$ is positive then 
\begin{equation*}
\left( \Psi _{-},(\sum \phi ^{\ast }\left( \left\vert v_{i}\right\rangle
\left\langle v_{j}\right\vert \right) \otimes e_{ij})\Psi _{-}\right) =-2{%
Re}\lambda _{k}\overline{\lambda _{l}}<0.
\end{equation*}%
Also if $2{Re}\lambda _{k}\overline{\lambda _{l}}$ is negative then%
\begin{equation*}
\left( \Psi _{+},(\sum \phi ^{\ast }\left( \left\vert v_{i}\right\rangle
\left\langle v_{j}\right\vert \right) \otimes e_{ij})\Psi _{+}\right) =2{Re%
}\lambda _{k}\overline{\lambda _{l}}<0.
\end{equation*}%
This means that $\phi ^{\ast }$ is non-CP.

\vskip 1cm

The above example can be readily generalized (cf Example 2). Namely, a normal state $\omega$ on $\cB(\cH \otimes \cK)$
can be written as
$$\omega(a \otimes b) = Tr_{\cH \otimes \cK} \varrho \  a\otimes b, $$
where $\varrho$  is the corresponding density matrix. But, the spectral representation of $\varrho$ implies
$$\omega(a \otimes b) = Tr_{\cH \otimes \cK} ( \sum_i\lambda_i P_{x_i} a \otimes b) = \sum_i \lambda_i \omega_{x_i}(a \otimes b).$$
Consequently, the entanglement mapping $\phi^{\ast}_{\omega}$ associated with the state $\omega$ would have the form
$$\phi^{\ast}_{\omega} = \sum_i \lambda_i \phi^{\ast}_{\omega_{x_i}} = \sum_i Tr_{\cH}(a \otimes I) P_{x_i} = Tr_{\cH} (a \otimes I) \varrho,$$
where the second equality follows from (\ref{stanczysty}).
Obviously, the analysis of CP and co-CP for $\phi^{\ast}_{\omega}$ is, in general, much more complicated.

Finally, to get a better understanding of the difference between CP and co-CP given in Example 3, let us consider the very particular case of this example.

\vskip 1cm

\textbf{Example 4. }
Let in Example 3, $\cH$ and $\cK$  be three dimensional Hilbert spaces. Further, put in the place of $x$  vectors giving maximally entangled pure states, i.e.

\begin{equation*}
x_1 = {\frac{1}{\sqrt{3}}} (e_1\otimes f_2 - e_2\otimes f_3 - e_3\otimes f_1)
\end{equation*}
and
\begin{equation*}
x_2 = {\frac{1}{\sqrt{3}}} (e_1\otimes f_1 + e_2\otimes f_2 + e_3\otimes f_3)
\end{equation*}

where $\{e_i\}_1^3$ ($\{f_i \}_1^3$) is a CONS in $\cH$ (in $\cK$ respectively). Easy calculations, which are left to the  reader, lead to the following maps
\begin{equation}
\label{dodat}
[a_{ij}]_{i,j=1}^3 \mapsto {\frac{1}{3}}
\left(
\begin{array}{ccc}
a_{33} & - a_{13} & a_{23}\\
-a_{31} & a_{11} & - a_{21} \\
a_{32} & - a_{12} & a_{22}\\
\end{array}  
\right)  
\end{equation}
for $x_1$, and

\begin{equation}
[a_{ij}]_{i,j=1}^3 \mapsto {\frac{1}{3}} ([a_{ij}]_{i,j=1}^3)^t.
\end{equation}
for $x_2$. Here, $([a_{ij}]_{i,j=1}^3)^t$ stands for the transposed map.
Clearly, transposition is not even 2-positive, so not CP. Now, non-CP observed of Example 3 should be well understood. The maps (\ref{dodat}) will be useful in the last Section.

\section{Tomita's scheme for partial transposition (see
\cite{II}, \cite{Majppt})}

Let $\mathcal{H}$ be a (separable) Hilbert space. Using an invertible density
matrix $\rho $ we can define a faithful state $\omega $ on $\mathcal{B}%
\left( \mathcal{H}\right) $ as $\omega \left( a\right) = Tr\rho a$ for $a\in 
\mathcal{B}\left( \mathcal{H}\right) $. Let us consider the GNS triple $%
\left( \mathcal{H}_{\pi },\pi ,\Omega \right) $ associated with $\left( 
\mathcal{B}\left( \mathcal{H}\right) \text{, }\omega \right) $. Such triple
is given by:

\begin{itemize}
\item GNS Hilbert space: $\mathcal{H}_{\pi }=\overline{\left\{ a\Omega \text{
};a\in \mathcal{B}\left( \mathcal{H}\right) \right\} }^{\left( \cdot ,\cdot
\right) }$ with $\left( a,b\right) =Tra^{\ast }b$ for $a,b\in \mathcal{B}%
\left( \mathcal{H}\right) .$

\item cyclic vector: $\Omega =\rho ^{1/2}$ .

\item representation: $\pi \left( a\right) \Omega =a\Omega .$
\end{itemize}

In the considered GNS representation, the modular conjugation $J_{m}$ is just the hermitian involution $%
J_{m}a\rho ^{1/2}=\rho ^{1/2}a^{\ast }$, and the modular operator $\Delta $
is equal to the map $\rho \cdot \rho ^{-1}$. However,
some remarks are necessary here. As we have assumed that $\cH$ is a separable Hilbert space then $\rho^{-1}$ is, in general, an unbounded operator. Hence, the domain of $\Delta$ should be described. To this end we note that: i) $\{A\rho ^{1/2}; A \in \cB(\cH) \}$ is a dense subset in the set of all Hilbert-Schmidt operators $\cF_{HS}(\cH)$
on the Hilbert space $\cH$, ii) $\alpha_t(\sigma) = \rho^{it} \sigma \rho^{-it}$ is an one parameter group of automorphisms on $\cF_{HS}(\cH)$. 
so, there exists (cf \cite{BR}) the set of entire analytic elements  $\cF_{HS}^0(\cH)$ of $\alpha_t(\cdot)$.
Thus $\Delta \sigma = \alpha_t(\sigma)|_{t = - i} = \rho \sigma \rho^{-1}$ is well defined for $\sigma \in
\cF_{HS}^0(\cH)$. In particular, the polar decomposition of Tomita's operator (cf. \cite{Takesaki}) is also well defined
\begin{equation}
SA\Omega = A^* \Omega = J_m \Delta^{1/2} A \Omega
\end{equation}
Note, that $ \{ A \Omega; A \in \cB(\cH) \} \subseteq D(\Delta^{1/2})$, $D(\cdot)$ stands for the domain.
In order to discuss the
transposition on $\pi \left( \mathcal{B}\left( \mathcal{H}\right) \right) $
we introduce the following two conjugations: $J_{c}$ on $\mathcal{H}$ and $J$
on $\mathcal{H}_{\pi }.$ Thanks to the faithfulness of $\omega $ the
eigenvectors $\left\{ e_{i}\right\} $ of $\rho $ form an orthogonal basis in 
$\mathcal{H}$. Hence we can define%
\begin{equation}
J_{c}x=\sum_{i}\overline{\left\langle e_{i},x\right\rangle }e_{i}
\end{equation}%
for every $x\in \mathcal{H}$. Due to the fact that $\left\{
E_{ij}=\left\vert e_{i}\right\rangle \left\langle e_{j}\right\vert \right\} $
form an orthogonal basis in $\mathcal{H}_{\pi }$ we can also define a
conjugation $J$ on $\mathcal{H}_{\pi }$%
\begin{equation}
Ja\Omega =\sum_{i}\overline{\left( E_{ij},a\Omega \right) }E_{ij}
\end{equation}%
with $J\Omega =\Omega $.

Following the construction presented in \cite{II} and \cite{Majppt} let us define a transposition on $\mathcal{B}%
\left( \mathcal{H}\right) $ as the map $a\in \mathcal{B}\left( \mathcal{H}%
\right) \mapsto a^{t}\equiv J_{c}a^{\ast }J_{c}$. By $\tau _{0}$ we will
denote the map induced on $\mathcal{H}_{\pi }$ by the transposition, i.e.%
\begin{equation*}
\tau _{0}a\Omega =a^{t}\Omega .
\end{equation*}%
Here are the main properties of $\tau _{0}$:

\begin{proposition}
(cf \cite{II}) 
\label{28}
(1) Let $a\in \mathcal{B}\left( \mathcal{H}\right) $ and $\xi \in 
\mathcal{H}_{\pi }$. Then%
\begin{equation}
a^{t}\xi =Ja^{\ast }J\xi .
\end{equation}%
(2) The map $\tau _{0}$ has a polar decomposition, i.e.%
\begin{equation}
\tau _{0}=U\Delta ^{1/2}
\end{equation}%
where $U$ is an unitary operator on $\mathcal{H}_{\pi }$ defined by $%
U=\sum_{ij}\left\vert E_{ij}\right) \left( E_{ji}\right\vert .$
\end{proposition}

where the sum defining the operator $U$ is understood in the weak operator topology.

In the above setting we can introduce the natural cone $\mathcal{P}$ (cf \cite{Araki}, \cite{Co})
associated with $\left( \pi \left( \mathcal{B}\left( \mathcal{H}\right)
\right) ,\Omega \right) $:%
\begin{equation*}
P=\overline{\left\{ \Delta ^{1/4}a\Omega :a\geq 0,a\in \pi \left( \mathcal{B}%
\left( \mathcal{H}\right) \right) \right\} }^{\left( \cdot ,\cdot \right) }.
\end{equation*}%
The relationship between the Tomita-Takesaki
scheme and transposition has the following form:

\begin{proposition}
\label{3.7}
(see \cite{II}) Let $\xi \mapsto \omega _{\xi }$ be the homeomorphism between the
natural cone $\mathcal{P}$ and the set of normal states on $\pi \left( 
\mathcal{B}\left( \mathcal{H}\right) \right) $, such that%
\begin{equation*}
\omega _{\xi }\left( a\right) =\left( \xi ,a\xi \right) ,\text{ }a\in 
\mathcal{B}\left( \mathcal{H}\right) .
\end{equation*}%
For every state $\omega $ define $\omega ^{\tau }\left( a\right) =\omega
\left( a^{t}\right) $. If $\xi \in \mathcal{P}$ then the unique vector in $%
\mathcal{P}$ mapped into the state $\omega _{\xi }^{\tau }$ by the
homeomorphism described above, is equal to $U\xi ,i.e.$%
\begin{equation*}
\omega _{\xi }^{\tau }\left( a\right) =\left( U\xi ,aU\xi \right) ,\text{ }%
a\in \mathcal{B}\left( \mathcal{H}\right) .
\end{equation*}
\end{proposition}

\section{PPT states, a Hilbert space approach}

Let $\mathcal{H}_{A}$ and $\mathcal{H}_{B}$ be  finite dimensional Hilbert spaces. 
We want to emphasize that finite dimensionality of Hilbert spaces is assumed only in the proof of Theorem \ref{TheoremPPT}.
More precisely, due to technical questions concerning the domain of modular operator $\Delta$ we were able to prove this Theorem only for finite dimensional case (see \cite{II}). On the other hand, we emphasize that the description of compound system based on Tomita's approach is very general.  It relies on the construction of tensor product of standard forms of von Neumann algebras and this description can be done in very general way (so infinite dimensional case is included, cf \cite{Cu}).

Again let us consider a composite system $A+B$. Suppose that the subsystem $A$ is
described by $\mathcal{A=}$ $\mathcal{B}\left( \mathcal{H}_{A}\right) $ and is 
equipped with a faithful state $\omega _{A}$ given by an invertible density
matrix $\rho _{A}$ as $\omega _{A}\left( a\right) \equiv Tr\rho _{A}a$.
Similarly, let $\mathcal{B=}$ $\mathcal{B}\left( \mathcal{H}_{B}\right) $ define
the subsystem $B$, $\rho _{B}$ be an invertible density matrix in $%
\mathcal{B}\left( \mathcal{H}_{B}\right) $ and $\omega _{B}$ be a state on $%
\mathcal{B}$ such that $\omega _{B}\left( b\right) \equiv Tr\rho _{B}b$ for $%
b\in \mathcal{B}$. By $\left( \mathcal{K},\pi ,\Omega \right) $, $\left( 
\mathcal{K}_{A},\pi _{A},\Omega _{A}\right) $ and $\left( \mathcal{K}%
_{B},\pi _{B},\Omega _{B}\right) $ we denote the GNS representations of $%
\left( \mathcal{A}\otimes \mathcal{B}\text{, }\omega _{A}\otimes \omega
_{B}\right) $, $\left( \mathcal{A}\text{, }\omega _{A}\right) $ and $\left( 
\mathcal{B}\text{, }\omega _{B}\right) $ respectively. Then the triple $%
\left( \mathcal{K},\pi ,\Omega \right) $ can be given by the following
identifications (cf \cite{Cu}, \cite{MajOSID}):%
\begin{equation*}
\mathcal{K}=\mathcal{K}_{A}\otimes \mathcal{K}_{B}\text{, }\pi =\pi
_{A}\otimes \pi _{B}\text{, }\Omega =\Omega _{A}\otimes \Omega _{B}.
\end{equation*}%
With these identifications we have%
\begin{equation*}
J_{m}=J_{A}\otimes J_{B}\text{, }\Delta =\Delta _{A}\otimes \Delta _{B}
\end{equation*}%
where $J_{m}$, $J_{A}$, $J_{B}$ are modular conjugations and $\Delta $, $%
\Delta _{A}$, $\Delta _{B}$ are modular operators for $\left( \pi \left( 
\mathcal{A}\otimes \mathcal{B}\right) ^{\prime \prime }\text{, }\Omega
\right) ,\left( \pi _{A}\left( \mathcal{A}\right) ^{\prime \prime }\text{, }%
\Omega _{A}\right) $, $\left( \pi _{B}\left( \mathcal{B}\right) ^{\prime
\prime }\text{, }\Omega _{B}\right) $ respectively. Due to the finite
dimensionality of the corresponding Hilbert spaces, just to simplify our notation, we will identify $\pi
_{A}\left( \mathcal{A}\right) ^{\prime \prime }$ and $\pi _{A}\left( \mathcal{A}%
\right) $, etc. Moreover we will also write $a\Omega _{A}$ and $b\Omega _{B}$
instead of $\pi _{A}\left( a\right) \Omega _{A}$ and $\pi _{B}\left(
b\right) \Omega _{B}$ for $a\in \mathcal{A}$, $b\in \mathcal{B}$ when no
confusion can arise. Furthermore we denote the finite dimension of $\mathcal{H}_{B}$
by $n$. Thus $\mathcal{B}\left( \mathcal{H}_{B}\right) \equiv \mathcal{B}%
\left( \mathbb{C}^{n}\right) \equiv M_{n}\left( \mathbb{C}\right) $. To put
some emphasis on the dimensionality of the "reference" subsystem $B$,  we denote by $%
\mathcal{P}_{n}$ the natural cone for $\left( M_{n}^{\pi }\left( 
\mathcal{A}\right) ,\omega _{A}\otimes \omega _{0}\right) $, where $\pi
\left( \mathcal{A}\otimes M_{n}\left( \mathbb{C}\right) \right) $ is denoted
by $M_{n}^{\pi }\left( \mathcal{A}\right) $ and $\omega _{0}$ is a faithful
state on $M_{n}\left( \mathbb{C}\right) $.

In order to characterize the set of PPT states we need the notion
of the "transposed cone" $\mathcal{P}_{n}^{\tau }=\left( I\otimes U\right) 
\mathcal{P}_{n}$, where $\tau $ is the transposition on $M_{n}\left( \mathbb{C}%
\right) $ and \ $U$ is the unitary operator given in Proposition \ref{28} with the
eigenvectors of density matrix $\rho _{0}$ corresponding to $\omega _{0}$.

Then the construction of $\mathcal{P}_{n}$ and $\mathcal{P}_{n}^{\tau }$ may be
realized as follows:%
\begin{equation*}
\mathcal{P}_{n}=\overline{\left\{ \Delta ^{1/4}\left[ a_{ij}\right] \Omega :%
\left[ a_{ij}\right] \in M_{n}^{\pi }\left( \mathcal{A}\right) ^{+}\right\} }%
,
\end{equation*}%
\begin{equation*}
\mathcal{P}_{n}^{\tau }=\overline{\left\{ \Delta ^{1/4}\left[ a_{ji}\right]
\Omega :\left[ a_{ij}\right] \in M_{n}^{\pi }\left( \mathcal{A}\right)
^{+}\right\} }.
\end{equation*}%

Consequently, we arrived to

\begin{theorem}
\label{TheoremPPT}
(see \cite{II}) In the finite dimensional case 
\begin{equation*}
\mathcal{P}_{n}^{\tau }\cap \mathcal{P}_{n}=\left\{ \Delta ^{1/4}\left[
a_{ij}\right] \Omega :\left[ a_{ij}\right] \geq 0,\left[ a_{ji}\right] \geq
0\right\} .
\end{equation*}
\end{theorem}

\begin{corollary}
\label{14}
(1) There is one to one correspondence between the set of PPT states and $%
\mathcal{P}_{n}^{\tau }\cap \mathcal{P}_{n}.$

(2) There is one to one correspondence between the set of separable states
and $\mathcal{P}_{A}\otimes \mathcal{P}_{B}$ (cf \cite{MajOSID}).
\end{corollary}

\begin{remark}
The correspondence given in Corollary \ref{14}.2 holds for a general case. Thus, the above characterization is applicable to a true quantum system.
\end{remark}

We wish to close this Section with the following remark. Also, here, in the Hilbert space approach we met many ``cones'': ${\mathcal P}_n$, ${\mathcal P}_n^{\tau}$, ${\mathcal P}_{\cA} \otimes {\mathcal P}_{M_n}$. All these cones, as we have seen, play the crucial role in the description of important classes of states: all states, PPT states, and separable states respectively . This should be considered as another manifestation of ``mysterious behavior'' of tensor products (see Section 1).

\section{Equivalence between two types of characterization of PPT states}

In this Section, we wish to discuss the relation between the Hilbert space description of PPT states and
B-O characterization. Firstly, we note that Tomita's approach  
leads to the following representation of the compound state $\omega $: 
\begin{equation*}
\omega \left( \sum_{i}a_{i}\otimes b_{i}\right) 
= \sum_i (\xi, a_i \otimes b_i \xi)
=\sum_{i}\varphi _{\xi
,a_{i}}\left( b_{i}\right)
\end{equation*}%
where $\varphi _{\xi ,a_{i}}\left( b_{i}\right) \equiv \left( \xi ,\left(
a_{i}\otimes b_{i}\right) \xi \right)$ and $\xi \in \cP_n.$ 
We have used here the well known result from Tomita-Takesaki theory saying that for any normal state $\omega$
on a von Neumann algebra with cycling and separating vector $\Omega$ there is \textbf{unique} vector $\xi$ in the natural cone $\cP_n$ such that $\omega (a) = (\xi, a \xi)$.

Let us observe that for $a \in \cA$, $a \geq0$, and any $b \in \cB$
\begin{eqnarray*}
\omega(a \otimes b^t)
& =& (\xi, a \otimes b^t \xi) \\
&=& Tr_{\cK} |\xi><\xi| a^{\frac{1}{2}} \otimes 1 \cdot 
a^{\frac{1}{2}} \otimes 1 \cdot 1 \otimes b^t\\
&=& Tr_{\cK_A} Tr_{\cK_B} a^{\frac{1}{2}} \otimes 1 \cdot|\xi><\xi| \cdot a^{\frac{1}{2}} \otimes 1 \cdot 1 \otimes b^t\\
&=& Tr_{\cK_B} \Bigl(Tr_{\cK_A} a^{\frac{1}{2}} \otimes 1 \cdot|\xi><\xi| \cdot a^{\frac{1}{2}} \otimes 1\Bigr) \cdot  b^t\\
&=& Tr_{\cK_B} ({\rho}_{A \xi} b^t) = (\chi_{\rho, \xi}, b^t \chi_{\rho, \xi})\\
&=& (U\chi_{\rho, \xi}, b U\chi_{\rho, \xi}) = (\chi_{\rho, \xi}, UbU\chi_{\rho, \xi})\\
&=& Tr_{\cK_B} \rho_{A,\xi} UbU = \omega(a \otimes UbU)
\end{eqnarray*}
where $\chi_{\rho, \xi}$ is Tomita's representation of $Tr_{\cK_B}( \rho_{A, \xi} \cdot)$ and we have used the notation given in Section 5, Proposition \ref{3.7} and that the fact the partial trace $Tr_{\cK_A} (\cdot)$ is well defined conditional expectation.

As $\omega(a \otimes b)$ is linear in $a$, and any $a$ can be written as a sum of four positive elements (Jordan decomposition)
the previous result can be extended to
\begin{equation}
\label{U}
\omega(a \otimes b^t) = \omega(a \otimes UbU)
\end{equation}
for any $a \in \cA$ and $b \in \cB$.

Now we are in position to compare the strategy given by Lemma \ref{drugi lemat} and B-O approach with the Hilbert space description of PPT states. Firstly we note (cf Lemma \ref{drugi lemat}) that maps $\varphi _{\xi , \cdot}\left( \cdot \right)$
can be considered as 
\begin{equation}
\cB(\cK_A) \ni a \mapsto \varphi _{\xi ,a}\left( \cdot \right) \in \cB(\cK_B)_*
\end{equation}

Secondly, note that the positivity used in Lemma \ref{drugi lemat}(2) implies
\begin{eqnarray*}
0 &\leq &\omega \left( \sum_{i,j}a_{i}^{\ast }a_{j}\otimes b_{i}^{\ast
}b_{j}\right) \\
&=&\sum_{i,j}\varphi _{\xi ,a_{i}^{\ast }a_{j}}\left( b_{i}^{\ast
}b_{j}\right) = \sum_{i,j}\left( \xi ,\left( a_{i}^{\ast }a_{j}\otimes b_{i}^{\ast
}b_{j}\right) \xi \right). 
\end{eqnarray*}%

By using the same vector $\xi \in \mathcal{P}_{n}$ let us define
 $\omega ^{\tau }\in \left( \mathcal{A}\otimes 
\mathcal{B}\right)_* $

\begin{equation*}
\omega ^{\tau }\left( \sum_{i}a_{i}\otimes b_{i}\right) \equiv
\sum_{i}\varphi _{\xi ,a_{i}}^{\tau }\left( b_{i}\right)
\end{equation*}%
where $\varphi _{\xi ,a_{i}}^{\tau }\left( b_{i}\right) \equiv \left( \xi
,\left( a_{i}\otimes b_{i}^{t}\right) \xi \right) .$ 
The positivity used in Lemma \ref{drugi lemat}(1) implies
\begin{eqnarray*}
\omega ^{\tau }\left( \sum_{i,j}a_{i}^{\ast }a_{j}\otimes b_{i}^{\ast
}b_{j}\right) &=&\sum_{i,j}\varphi _{\xi ,a_{i}^{\ast }a_{j}}^{\tau }\left(
b_{i}^{\ast }b_{j}\right) \\
&=& \sum_{i,j}\left( \xi ,\left( a_{i}^{\ast }a_{j}\otimes \left( b_{i}^{\ast
}b_{j}\right) ^{t}\right) \xi \right) \\
&=& \sum_{i,j}\left( \xi ,\left( a_{i}^{\ast }a_{j}\otimes b_{j}^{t}\left( b_{i}^{\ast
}\right) ^{t}\right) \xi \right)\\
&=& \sum_{i,j}\left(I\otimes U \xi ,\left( a_{i}^{\ast }a_{j}\otimes b_{j}\left( b_{i}^{\ast
}\right) \right)I\otimes U \xi \right) \geq 0,\\
\end{eqnarray*}%

where in the last equality we have used (\ref{U}).
Hence, CP and co-CP of entangling mapping is equivalent to
 $\xi \in 
\mathcal{P}_{n}^{\tau }\cap \mathcal{P}_{n}$. Consequently, we conclude that

\begin{theorem}
The description of PPT states by $\mathcal{P}_{n}^{\tau }\cap \mathcal{P}%
_{n} $ can be recognized as the dual description of PPT states by $\mathcal{E%
}/\mathcal{E}_{q}.$
\end{theorem}

In the base of the above equivalence of two types of description of PPT
states we may discuss the effectiveness of such characterizations from
different points of view. This will be the topic of next Sections. We will start with an analysis of decomposable maps (cf \cite{Mst}).

\section{ On decomposable maps}
In \cite{St3}
St{\o}rmer gave the following characterization of decomposable maps:
\begin{theorem}{\rm(\cite{St3})}
\label{Stormer}
Let $\phi: \cA \to B(\cH)$ be a positive map.
A map $\phi$ is decomposable if and only if for all $n \in \bN$ whenever $[x_{ij}]$ and $[x_{ji}]$ belong to $M_n(\cA)^+$ then $[\phi(x_{ij})]\in M_n(B(\cH))^+$.
\end{theorem}

As our aim is to discuss effectiveness of description of PPT states given in Section 5, we again assume finite dimensionality of Hilbert space $\cH$. Further, 
recall (see Criterion \ref{kryt1}) that 
 the positivity of the matrix $[\phi(x_{ij})]$ (with operator entries!) is equivalent to
\begin{equation}
\label{dwa}
\sum_{ij} y^*_i \phi(x_{ij}) y_j \geq 0
\end{equation}
where $\{ y_i\}$ are arbitrary elements of $B(\cH)$.
Furthermore, any positive matrix $[x_{ij}]$ can be written as (cf \cite{Tak})

\begin{equation}
\label{hoho}
[x_{ij}] = \sum_k [(v^{(k)}_{i})^* v^{(k)}_{j}]
\end{equation}
Hence, applying  condition (\ref{dwa}) to matrices of the form $[a^*_i a_j]$ with the choice of $y_i$ such that all $y_i = 0$ except for $i_0$ and $j_0$, then changing the numeration in such way that $y_{i_0} = y_1$ and $y_{j_0} = y_2$ we arrive to
study the positivity of the following matrix
\begin{equation}
\label{jeden}
\left(
\begin{array}{cc}
a^*_1a_1 & a^*_1a_2 \\
a^*_2 a_1 & a^*_2 a_2 \\
\end{array}  
\right)  
\geq 0            
\end{equation}                
and its transposition.
On the other hand,  block matrix techniques leads to  necessary and sufficient  conditions for positivity of such matrices. Namely, let $A,B,C$ be $d \times d$ matrices. Then
\begin{lemma} {\rm (see \cite{XZ})} 
\label{Zhan}
\label{XZ}
\begin{equation}
\Big[
\begin{array}{cc}
A & B \\
B^* & C \\
\end{array}  
\Big]  
\geq 0            
\end{equation}                
if and only if $A\geq 0$, $C\geq 0$ and there exists a contraction $W$ such that $B = A^{\frac{1}{2}} W C^{\frac{1}{2}}$.
\end{lemma}
Assume, if necessary, that $a_1$ and $a_2$ have inverses, otherwise $a_i^{-1}$ is understood to be generalized inverse of $a_i$. Then, application of Lemma \ref{XZ} to the St{\o}rmer condition leads to the following question:
When $|a_1|^{-1} a^*_2 a_1 |a_2|^{-1}$ is a contraction?
But an operator $T \in B(\cH)$ is a contraction if and only if $||T||\leq 1$ what is equivalent to
$||Tx||^2 \leq ||x||^2$. This can be written as
\begin{equation}
\label{contraction}
(x,T^*Tx) \leq (x,x) 
\end{equation} what is equivalent to
\begin{equation}
\label{contraction2}
T^*T\leq \bf 1
\end{equation}

Consequently, (\ref{contraction2}) and Zhan's lemma \ref{XZ} give (see also \cite{A1} and \cite{Ch2})
\begin{equation}
a^*_1a_2|a_1|^{-2} a_2^* a_1 \leq |a_2|^2
\end{equation}
Hence
\begin{equation}
\forall_f \quad (f,a^*_1a_2 (a_1^*a_1)^{-1} a^*_2a_1 f) \leq (f, a_2^* a_2 f)
\end{equation}
So, putting $f = a_1^{-1} g$ one gets
\begin{equation}
\forall_g \quad ||(a^*_1)^{-1} a^*_2 g|| \leq ||a_2 a_1^{-1} g||
\end{equation}
This means hyponormality of operators $(a_2 a_1^{-1})^*$ (cf. \cite{H}, and \cite{Stamp}).
But, as considered operators are defined on a finite dimensional Hilbert space, in particular, they are completely continuous.
Therefore, hyponormality of $(a_2 a_1^{-1})^*$ implies normality (see \cite{A}, \cite{B}, and \cite{Stamp}).

Consequently, $a_2 a^{-1}_1$ is a normal operator.
This means that there is a unitary operator $U$ (equivalently unitary matrix as finite dimensions are assumed) such that
\begin{equation}
U a_2 a_1^{-1} U^* = diag(\lambda_i)
\end{equation}
where $\lambda_i \in \bC$. This can be rewritten as
\begin{equation}
a_2a_1^{-1} = \sum_i \lambda_i Q_i
\end{equation}
where $\lambda_i \in \bC$ and $\{Q_i\}$ is the resolution of identity. Hence,
putting 
\newline
$Q_i \equiv |e_i><e_i|$ where $\{ e_i \}$ is a CONS in the Hilbert space $\cH$ on which operators $\{ a_i \}$ act and  
defining rank one operators
$|f><g|z \equiv (g,z) |f>$, one gets
\begin{equation}
\label{lala}
a_2 = \sum_i \lambda_i |e_i><a^*_1e_i|
\end{equation}

Thus we proved:
\begin{proposition}
\label{proposition}
For any matrix
$\left(
\begin{array}{cc}
a^*_1a_1 & a^*_1a_2 \\
a^*_2 a_1 & a^*_2 a_2 \\
\end{array}  
\right)$  
satisfying the St{\o}rmer condition, $a_2$ is of the form  (\ref{lala}).
\end{proposition}        

\begin{remark}
Using the Ando-Choi inequality (see \cite{A1}, \cite{Ch2}) one gets analogous formula for $a_1$ in terms of $a_2$.
\end{remark}
As a next step we note that (\ref{lala}) and St{\o}rmer condition lead to the following form of the matrix $\left(
\begin{array}{cc}
a^*_1a_1 & a^*_1a_2 \\
a^*_2 a_1 & a^*_2 a_2 \\
\end{array}  
\right)$: 

\begin{equation}
\left(
\begin{array}{cc}
a^*_1a_1 & a^*_1a_2 \\
a^*_2 a_1 & a^*_2 a_2 \\
\end{array}  
\right) = \sum_i \left(
\begin{array}{cc}
1 & \lambda_i \\
\bar{\lambda_i} & |\lambda_i|^2 \\
\end{array}  
\right) \dot 
{\Big(
\begin{array}{cc}
|a_1^*e_i><a_1^*e_i| & 0 \\
0 & |a_1^*e_i><a_1^* e_i| \\
\end{array}  
\Big)}   
\end{equation}

To rewrite the above equality in a more compact form, let us denote the norm of the vector $|a_1^*e_i>$ by $\alpha_i$ and the normalized vector $\frac{1}{\alpha_i} |a_1^*e_i>$ by $\varphi_i$. Then

\begin{equation}
\label{baba1}
\left(
\begin{array}{cc}
a^*_1a_1 & a^*_1a_2 \\
a^*_2 a_1 & a^*_2 a_2 \\
\end{array}  
\right) = \sum_i \alpha_i^2 \left(
\begin{array}{cc}
1 & \lambda_i \\
\bar{\lambda_i} & |\lambda_i|^2 \\
\end{array}  
\right) \dot 
{\Big(
\begin{array}{cc}
|\varphi_i><\varphi_i| & 0 \\
0 & |\varphi_i><\varphi_i| \\
\end{array}  
\Big)}   
\end{equation}
or symbolically
\begin{equation}
\label{chocho1}
\left(
\begin{array}{cc}
a^*_1a_1 & a^*_1a_2 \\
a^*_2 a_1 & a^*_2 a_2 \\
\end{array}  
\right) = \sum_i \alpha_i^2  \cdot \Lambda_i \cdot R_i
\end{equation}

where $\Lambda_i$ are ``matrix'' coefficients while $R_i$ are ``matrix'' projectors (not mutually orthogonal!).
This leads to:
\begin{corollary}
\label{21}
  (\ref{chocho1}) implies ``separability'' for $[a_i^*a_j]$ satisfying the St{\o}rmer condition. 
  Namely, using the identification $M_2(\cB(\cH))\cong M_2(\bC)\otimes \cB(\cH)$ (cf discussion concerning equations (\ref{22}) and (\ref{23})) and noting that $(1 + |\lambda_i|^2)^{-\frac{1}{2}} \Lambda_i \equiv P_i$ is a projector one can write
  $$\left(
\begin{array}{cc}
a^*_1a_1 & a^*_1a_2 \\
a^*_2 a_1 & a^*_2 a_2 \\
\end{array}  
\right) = \sum_i \alpha_i^2 (1 + |\lambda_i|^2)(P_i \otimes \bf{1})(\bf{1}\otimes |\varphi_i><\varphi_i|).
$$
Hence
\begin{equation}
\left(
\begin{array}{cc}
a^*_1a_1 & a^*_1a_2 \\
a^*_2 a_1 & a^*_2 a_2 \\
\end{array}  
\right) \in M_2(\bC)^+\otimes \cB(\cH)^+.
\end{equation}
   Therefore, it is important to realize that non-triviality of St{\o}rmer condition follows from the fact
that when a positive matrix $[x_{ij}]$ {\rm (}$= \sum_k [(v^{(k)}_{i})^* v^{(k)}_{j}]${\rm)}  satisfies the St{\o}rmer condition some of its summand(s)
$[(v^{(k)}_{i})^* v^{(k)}_{j}]$  may not.
\end{corollary}

We end this Section with

\begin{remark} 
 Formula (\ref{baba1}) can serve as a part of recipe for producing PPT states and some non-decomposable maps on matrix algebras (see next Section).
\end{remark}

\section{Effectiveness of the description of PPT states}

Now we are able to discuss the question of effectiveness of the construction of $\cP_n \cap \cP^{\tau}_n$.
In other words we are interested in the following question: Can one provide a canonical form for a vector in $\cP_n \cap \cP^{\tau}_n$?

We begin with the remark that the structure of $\cP_n \cap \cP^{\tau}_n$ given by Theorem \ref{TheoremPPT} reflects the St{\o}rmer characterization of decomposable maps (see Theorem \ref{Stormer}). Hence the posed problem seems to be equivalent to the question whether the given characterization of decomposable maps is an effective one in the sense that we wish to know the canonical form of matrices $[a_{ij}]$ such that $[a_{ij}]\geq 0$ and $[a_{ji}]\geq 0$.

The important point to note here is the Tomiyama characterization of positive transpositions (see \cite{Tom2}).
Let $\mathcal A$ be a $C^*$-algebra. The transposition $\tau$ on the set of matrices $[a_{ij}]$ with $a_{ij} \in \cA$ is a positive map if and only if 
$\mathcal A$ is abelian. 
This result suggests that the condition $f \in \cP_n \cap \cP^{\tau}_n$  reflects a kind of ``local commutativity''.

Let us elaborate briefly this point.
Firstly, we note : $\Lambda_i$ (see formula \ref{chocho1}) is a matrix with complex entries and the transposition on such matrices is a positive map.
Secondly, $R_i$ is a matrix with operator entries but this matrix is diagonal. Thus, for transposition, $R_i$ is a fixed point. Furthermore, $\Lambda_i$ commutes with $R_i$. 
We emphasize that all these remarks stem from (\ref{baba1}), (\ref{chocho1}) - so this is a ``local'' property as we singled out two indices only. Nevertheless we can conclude : any summand of a positive matrix $[x_{ij}]$ in (\ref{hoho}) satisfying the St{\o}rmer condition has ``local-commutativity''  which guarantees the nice behavior (positivity) of the transposition. But not every summand in (\ref{hoho}) has this property (see Corollary \ref{21})!

Finally, we are able to discuss the question of
effectiveness of the description of PPT states. To this end we recall that Theorem \ref{TheoremPPT} says: PPT states are characterized (uniquely) by vectors of the form $[a_{ij}] \Omega = \sum_k [(a_{i}^{(k)})^*a_{j}^{(k)})] \Omega$ with $[a_{ji}]\geq0$, where the last equality follows from (\ref{hoho}). However, we would like to note here the important point (cf the discussion following (\ref{chocho1}) ): some summands $[(a_{i}^{(k)})^*a_{j}^{(k)})] \Omega$ may not be in $\cP_n\cap \cP_n^{\tau}.$  Consequently, some vectors in the subcone $\cP_n\cap \cP_n^{\tau}$ which represent non-trivial (that is non-separable) PPT states can be obtained as a convex hull of vectors in such way that some summand(s) is (are) not necessarily in this subcone.
This can be expected as $\cP_n \cap \cP_n^{\tau}$ is a convex set which could be ``far'' from being a simplex. In other words, a convex decomposition of a vector in $\cP_n \cap \cP_n^{\tau}$ is far from being the unique one.
Concluding, the presented arguments suggest that the universally effective prescription for a vector representing PPT state is not available.
However, the above discussion provides some recipe for  construction of concrete vectors in $\cP_n \cap \cP^{\tau}_n$.

\section{ Measures of entanglement.}

 In \cite{M2} and \cite{M3} using the $C^*$-algebraic approach to Quantum theory,  we have introduced  the degree of quantum correlations. The basic idea is to describe how  a given quantum system is close to the ``classical'' world.
We wish to repeat this idea but now in the context of Hilbert spaces (cf \cite{MajTok}). For that purpose we will employ the geometry of Hilbert spaces.
\begin{definition}
Let $\xi$ be a vector in the natural cone $\cP$ corresponding to a normal state  of a composite system $A+B$ (cf Proposition \ref{3.7}). Then
\begin{enumerate}
\item{} Degree of entanglement (or {\it quantum correlations}) is given by:
\begin{equation}
D_e(\xi) = inf_{\eta} \{ ||\xi - \eta||; \eta \in {\cP}_A \otimes {\cP}_B \}
\end{equation}
\item{} Degree of genuine entanglement (or {\it genuine quantum correlations}) is defined as
\begin{equation}
D_{ge}(\xi) = inf_{\eta} \{ ||\xi - \eta||; \eta \in \cP_n \cap \cP^{\tau}_n \}
\end{equation}
\end{enumerate}
\end{definition}

\vskip 1cm

We will briefly discuss the geometric idea behind this definitions. The key to the argument is the concept of convexity (in Hilbert spaces).
Namely, we observe

\begin{enumerate}
\item $\cP \supset{\cP}_A \otimes {\cP}_B $ is a  convex subset,
\item $\cP \supset\cP_n \cap \cP^{\tau}_n$ is a  convex subset,
\item The theory of Hilbert spaces says: $\exists! \ \xi_0 \in {\cP}_A \otimes {\cP}_B$, such that $D_e(\xi) = || \xi - \xi_0||$,
\item Analogously, $\exists! \ \eta_0 \in {\cP} \cap {\cP}^{\tau}$, such that $D_{ge}(\xi) = || \xi - \eta_0||$.
\end{enumerate}

The important point to note here is that we used the well known property of convex subsets in a Hilbert space: a closed convex subset $W$ in a Hilbert space $\cH$ contains the unique vector with the smallest norm. This ensures the existence of vectors $\xi_0$ and $\eta_0$ introduced in 3. and 4. respectively.

It is expected that any well defined entanglement measure $D(\cdot)$ should, at least, satisfy the following requirements (see \cite{Hor3}, \cite{Ozawa}, \cite{Ved}, \cite{Vid}, and \cite{Wei}):
\begin{enumerate}
\item $D(\xi)\geq 0$,
\item $D(\xi) = 0$ if $\xi$ is not entangled,
\item $D(U_A\otimes U_B \xi) = D(\xi)$ where $U_A$ ($U_B$) are unitary operators representing local symmetry for subsystem A (B respectively),
\item convexity, i.e. $\sum_i \alpha_i D(\xi_i) \geq D(\sum_i \alpha_i \xi_i)$,
\item continuity.
\end{enumerate}

Clearly, $D_e$ satisfies all above listed requirements, while for $D_{ge}$ 1-2 and 4-5 hold.
Note that the failure of 3 for $D_{ge}$ is not surprising. Namely, recall that for any automorphism
on a von Neumann algebra in the standard form there exists the unique unitary operator on the Hilbert space
which leaves the natural cone globally invariant (again this is a result of Tomita-Takesaki theory, see also Section 4).Thus unitary operators appearing in 3 can describe  a local symmetry. On the other hand it is hard to expect that the set of PPT states has such general symmetry.

Another desirable property of degree (measure) of entanglement would be the monotonicity with respect to arbitrary nonselective operations (see \cite{Ved2}). Here, nonselective operations are understood as CP maps on the set of observables.
If such a map $\psi$ leaves  the selected vector state $\omega_{\Omega}$ invariant (in the GNS construction $\Omega$ is interpreted either as the vacuum (field theoretic interpretation) or as equilibrium (statistical interpretation); so this assumption is natural) then, by the generalized Schwarz-Kadison inequality, $\psi$ induces the contraction $\hat{\psi}$ on the GNS space. Consequently such condition in our framework is obviously satisfied provided that $\hat{\psi}$ leaves $\cP_1 \otimes \cP_2$ globally invariant.

We end this review of properties of entanglement measures with the remark that the idea of measuring entanglement of vectors in terms of their distance to separable vectors appeared in the papers cited in this Section (see also \cite{Ar}). \textbf{BUT} our approach is carried out in a very different setting and our concept of degree of entanglement stems from the definition given in \cite{M3}. In particular, Ozawa arguments on Hilbert-Schmidt distance are not applicable here (cf \cite{Ozawa}).

To illustrate our measures of entanglement we present the example which could be considered as a continuation of Example 4 and it is based on a modification of Kadison-Ringrose arguments (cf \cite{KR}) with Tomita-Takesaki theory (cf \cite{BR}):

\vskip 1cm

\textbf{Example 5}
Let $\{ e_1, e_2, e_3 \}$ ($\{f_1, f_2, f_3 \}$) be an orthonormal basis in the three dimensional Hilbert space $\cH$ ($\cK$ respectively). By $P$ we denote the following rank one orthogonal projector
$$ P = \frac{1}{3} |e_1\otimes f_1 + e_2 \otimes f_2 + e_3 \otimes f_2>< e_1\otimes f_1 + e_2 \otimes f_2 + e_3 \otimes f_2| \equiv |x_2><x_2| \in \cB(\cH \otimes \cK)^+ 
$$
Let $S$ be an operator of the form
$$S = \sum_{i=1}^{k<\infty} a_i \otimes b_i $$
where $a_i \in \cB(\cH)^+$ and $b_i \in \cB(\cK)^+$.
It can be shown (see \cite{KR}) that 
$$ ||P - S || \geq \frac{1}{6} $$
where $||\cdot||$ stands for the operator norm. 
Any separable state on  $\cB(\cH)\otimes \cB(\cK)$ can be expressed in the form
\begin{equation}
\varrho_0 = \sum_{i=1}^{l< \infty} \omega_{z_i}\otimes \omega_{y_i}
\end{equation}
where $z_1, ...,z_l \in \cH$, $y_1,...,y_l \in \cK$, and the vector state $\omega_z$ is defined as $\omega_z(a) \equiv (z, az)$.
Then, again, following Kadison-Ringrose exercise
one can show that
\begin{equation}
||\omega_{x_2} - \varrho_0|| \geq \frac{1}{6}
\end{equation}
On the other hand (see \cite{BR}), if $\xi$  (a vector $\eta) \in \cP$ defines the normal positive form  $\omega_{\xi}$ ($\omega_{\eta}$ respectively) then one has
$$ ||\xi - \eta||^2 \leq || \omega_{\xi} - \omega_{\eta}|| \leq ||\xi - \eta|||| \xi + \eta||$$
Consequently 
\begin{equation}
D_e(\varrho^{\frac{1}{4}}P\varrho^{\frac{1}{4}})\geq \frac{1}{12}.
\end{equation}
Obviously, $\varrho \equiv \varrho_{\cH} \otimes \varrho_{\cK}$, etc (cf Sections 4 and 5).
We end this example with a remark that the same arguments applied to $x^{\prime}_2 = \frac{1}{\sqrt{2}}(e_1\otimes f_1 + e_2 \otimes f_2)$ in 2D case lead to
\begin{equation}
D_e(\varrho^{\frac{1}{4}}P\varrho^{\frac{1}{4}})\geq \frac{1}{8}.
\end{equation}

\vspace {1cm}

Concluding this Section, for \textbf{an entangled (non-PPT state) we are able to find  the best approximation among separable states (PPT states, respectively)}. Moreover, this approach offers a classification of entanglement (genuine entanglement, respectively). 

\section{Final remarks}
In Section 1, we emphasized that we should deal with many cones. In other words, many types of ``positivity'' should be taken into account. Furthermore, Lemmas \ref{pierwszy lemat}
and \ref{drugi lemat} employ various ``orders'': (plain) positivity for the first lemma and CP for the second. On the other hand, we recall that there are many examples on non-CP maps.
Consequently, there appears natural question how to understand the difference between Lemmas \ref{pierwszy lemat} and \ref{drugi lemat} in this context. To  clarify these subtleties, in this Section we will argue that following the scheme offered by Lemma \ref{pierwszy lemat} one can use  (very) non-CP maps to describe states on $(\cA \odot \cB, \cA^+ \otimes \cB^+)$ which could have very strong correlations (cf \cite{Sudershan}).
To make our presentation as simple as possible we restrict ourselves to 3 dimensional case. We recall that 3 dimensional models are the simplest cases with non-decomposable maps (see \cite{Ch1}, \cite{W}, and \cite{Ch}). We begin with recalling 
Cho, Kye and Lee \cite{Cho} results. They studied the following family of maps $\bC^3 \to \bC^3$
\begin{equation}
\phi[a,b,c](x)= \psi[a,b,c](x) - x
\end{equation}

\begin{equation}
\psi[a,b,c](x_{ij})=
\left(
\begin{array}{ccc}
ax_{11}+ bx_{22} +c x_{33} & 0 &0 \\
0 & ax_{22}+bx_{33} + c x_{11} & 0 \\
0 & 0 & a x_{33} +b x_{11} + c x_{22}\\
\end{array}  
\right)
\end{equation}

The properties of these maps are collected in the following Theorem (see \cite{Cho})
\begin{theorem}
\label{CHO}
\begin{enumerate}
\item $\phi[2,0,\mu]$ for $\mu \geq 1$ are indecomposable
\item  $\phi[2,0,1]$ is atom
\item  $\phi[a,b,c]$ is positive if and only if $a\geq 1,\ a+b+c \geq 3,$ and  $ bc \geq(2 - a)^2$ if $1 \leq a\leq 2$
\item  $\phi[a,b,c]$ is CP f and only if $a\geq3$
\item  $\phi[a,b,c]$ is decomposable if and only if $a \geq 1$, $bc \geq ({\frac{3 -a}{2}})^2$ if $1 \leq a 
\leq 3$
\end{enumerate}
\end{theorem}

For some choices of the parameters $a,b,c$ one can arrive to maps very similar to that given by (\ref{dodat}) (see Section 3).
But, as Theorem \ref{CHO} is saying, there are many concrete very non-CP maps. We can use them and Lemma \ref{pierwszy lemat}
to produce very ``quantum'' functionals on compound systems - note that Lemma \ref{drugi lemat} always deals with CP maps and states (normalized positive functionals with respect to the cone $(\cA \otimes \cB)^+$!) on \Cs-algebraic tensor product.
Therefore, following Lemma \ref{pierwszy lemat}, we will define states $\omega$ by
\begin{equation}
\label{Kwantowy funcjonal}
\omega(a\otimes b)= Tr \phi(a)b^t
\end{equation}
where $\phi$ is any map described by Theorem \ref{CHO}.
Any $a,b \in \cB(\bC^3)$ can be written as
\begin{equation*}
a = \sum a_{ij} E_{ij},   \ b = \sum b_{kl} F_{kl}
\end{equation*}
where $a_{ij}, b_{kl} \in \bC$ while $E_{ij}$, $F_{kl}$ are basis in $\cB(\bC^3)$.
Hence
\begin{equation}
\label{53}
\sum_{ijkl} a_{ij} b_{kl} \omega(E_{ij} \otimes F_{kl}) = Tr \phi[abc](a) b^t = \sum_{ijkl} a_{ij} b_{kl} Tr \phi[abc](E_{ij}) F_{lk}
\end{equation}

But
\begin{equation}
\phi[abc](E_{ij}) = \psi[abc](E_{ij}) - E_{ij}
\end{equation}
Let  $i\neq j$. Then
\begin{equation}
\label{55}
\phi[abc](E_{ij}) = - E_{ij}
\end{equation}
To consider the case $i=j$ define
\begin{equation}
f_k(l) = a_{kl}
\end{equation}
where
\begin{equation}
[a_{kl}]=
\left(
\begin{array}{ccc}
a & b & c \\
c & a & b \\
b & c & a\\
\end{array}  
\right)
\end{equation}
Then
\begin{equation}
\label{58}
\phi[abc](E_{ii}) = -E_{ii} +
\left(
\begin{array}{ccc}
f_1(i) & 0 & 0 \\
0 & f_2(i) & 0 \\
0 & 0 & f_3(i)\\
\end{array}  
\right) 
\end{equation}

 Taking $a=2$, $b=0$, and $c>1$ , equations (\ref{53}), (\ref{55}), and (\ref{58}) give very ``quantum'' functionals, positive on the projective cone (so admitting negative values on $\C_{inj} \setminus \C_{pro}$) and showing another difference between Lemmas \ref{pierwszy lemat} and \ref{drugi lemat}.
Moreover, this shows how powerful ``machinery'' was proposed in Section 2 as well as gives another explanation of question studied in \cite{Sudershan}.
 
\section{Acknowledgments} 
We are grateful to V. P. Belavkin and M. Marciniak for fruitful discussions on positive maps.
W.A.M and T. M. would like to acknowledge
the supports of QBIC grant  and  W.A.M. acknowledges also the partial support of BW grant 5400-5-0089-8.

\end{document}